\documentclass{aa}  

\usepackage{natbib}
\bibpunct{(}{)}{;}{a}{}{,}
\usepackage{graphicx}
\usepackage{txfonts}
\DeclareRobustCommand{\uppartial}{\text{\rotatebox[origin=t]{20}{\scalebox{1}[1]{$\partial$}}}\hspace{-1pt}} 
\usepackage{geometry}
\usepackage{amsmath,amssymb}
\usepackage{bm}

\DeclareMathAlphabet{\mathbfsf}{\encodingdefault}{\sfdefault}{bx}{n}

\usepackage{subcaption}
\usepackage{minibox}
\usepackage{color}
\usepackage{upgreek}
\usepackage{pifont}
\usepackage{empheq}
\usepackage{physics}

\usepackage{pgf, tikz}
\usepackage{tikz-3dplot}
\usepackage{pgfplots}
\usetikzlibrary{calc}
\usetikzlibrary{shapes.misc}
\usetikzlibrary{shapes,snakes}
\usetikzlibrary{arrows}
\usetikzlibrary{automata,positioning}
\tikzset{
  load/.style   = {ultra thick,-latex},
  stress/.style = {-latex},
  dim/.style    = {latex-latex},
  axis/.style   = {-latex},
}
\tikzset{dimetric2/.style={
  x={(0.935cm,-0.118cm)},
  y={(0.354cm, 0.312cm)},
  z={(0.000cm, 0.943cm)},
}}
\usepackage{color}
\definecolor{auburn}{rgb}{0.43, 0.21, 0.1}
\definecolor{oxfordblue}{rgb}{0.0, 0.13, 0.28}
\definecolor{coolblack}{rgb}{0.0, 0.18, 0.39}
\definecolor{darkcerulean}{rgb}{0.03, 0.27, 0.49}
\definecolor{denim}{rgb}{0.08, 0.38, 0.74}
\definecolor{purpletaupe}{rgb}{0.31, 0.25, 0.3}
\definecolor{purpletaupemodif}{rgb}{0, 0, 0}
\definecolor{applegreen}{rgb}{0.55, 0.71, 0.0}
\definecolor{chaptercolor}{rgb}{0.2, 0.2, 0.2}
\definecolor{blue(ryb)}{rgb}{0.01, 0.28, 1.0}
\definecolor{ao}{rgb}{0.0, 0.0, 1.0}
\definecolor{electricultramarine}{rgb}{0.25, 0.0, 0.95}
\definecolor{electricpurple}{rgb}{0.75, 0.0, 1.0}
\definecolor{palatinatepurple}{rgb}{0.41, 0.16, 0.38}
\definecolor{aureolin}{rgb}{0.99, 0.93, 0.0}
\definecolor{fluorescentorange}{rgb}{1.0, 0.85, 0.0} 

\begin{document} 

   \title{Layered semi-convection and tides in giant planet interiors}

   \subtitle{II.  Tidal dissipation}

   \author{Q. Andr\'e\inst{1}
          \and S. Mathis\inst{1}\fnmsep\inst{2}
          \and A. J. Barker\inst{3}
          }
   \institute{AIM, CEA, CNRS, Universit\'e Paris-Saclay, Université Paris Diderot, Sorbonne Paris Cit\'e, F-91191 Gif-sur-Yvette, France\\
              \email{quentin.andre@cea.fr, stephane.mathis@cea.fr}
          \and
         	LESIA, Observatoire de Paris, CNRS UMR 8109, UPMC, Universit\'e Paris-Diderot, 5 place Jules Janssen, 92195 Meudon, France
          \and
             Department of Applied Mathematics, School of Mathematics, University of Leeds, Leeds, LS2 9JT, UK\\
             \email{a.j.barker@leeds.ac.uk}  
             }

   \date{Received XXXX; accepted YYYY}

  \abstract
   {{Recent Juno observations have suggested that the heavy elements in Jupiter could be diluted throughout a large fraction of its gaseous envelope, providing a stabilising compositional gradient over an extended region of the planet. This could trigger layered semi-convection, which, in the context of giant planets more generally,} may explain Saturn's luminosity excess and play a role in causing the abnormally large radii of some hot Jupiters. In giant planet interiors, {it could take the form of density staircases}, which are convective layers separated by thin stably stratified interfaces. In addition, the efficiency of tidal dissipation is known to depend strongly on the planetary internal structure.}
   {We aim to study the resulting tidal dissipation when internal waves are excited in a region of layered semi-convection by tidal gravitational forcing due to other bodies (such as moons in giant planet systems, or stars in hot Jupiter systems).}
   {{We adopt a local Cartesian model with a background layered density profile subjected to an imposed tidal forcing, and we compute the viscous and thermal dissipation rates numerically. We consider two sets of boundary conditions in the vertical direction: periodic boundaries, and impenetrable, stress-free boundaries, with periodic conditions in the horizontal {directions} in each case}. These models are appropriate for studying the forcing of {short-wavelength tidal} waves in part of a region of layered semi-convection, {and in an extended envelope containing layered semi-convection, respectively}.}
   {We find that the rates of tidal dissipation can be enhanced in a region of layered semi-convection compared to a uniformly convective medium, where the latter corresponds with the usual assumption adopted in giant planet interior models. In particular, a region of layered semi-convection possesses a richer set of resonances, allowing enhanced dissipation for a wider range of tidal frequencies. The details of these results significantly depend on the structural properties of the layered semi-convective regions.}
   {Layered semi-convection could contribute towards explaining the high tidal dissipation rates observed in Jupiter and Saturn, which have not yet been {fully} explained by theory. Further work is required to explore the efficiency of this mechanism in global models.}


   \maketitle
   
   \defcitealias{ABM2017}{Paper I}

\section{Introduction}\label{SEC:intro}
%
Based on astrometric measurements spanning more than a century \cite{LaineyEtal2009, LaineyEtal2012, LaineyEtal2017} found that the rates of tidal dissipation in Jupiter and Saturn {are} higher than previously thought {by one order of magnitude.} This has important astrophysical consequences since tidal interactions {are} a key mechanism {for driving} the rotational, orbital and thermal evolution of moons and planets (and also stars) over very long time-scales. 

Moreover, we know that this evolution, linked to the efficiency of tidal dissipation in celestial bodies, strongly depends on their internal structure{s} \citep[we refer the reader to the reviews by][and references therein]{MathisRemus2013,Ogilvie2014,Mathis2017}. {In this framework,} seismology has proven itself very useful for inferring the properties of, and understanding, the Earth's interior. However, for Jupiter and Saturn, there have been no clear detections of oscillations of their surfaces because the radial velocities of the excited modes have very small amplitudes \citep{GaulmeEtal2011}. Thus, the internal structures of giant planets remain poorly constrained. 

However, some progress has been made recently. By analysing the properties of density waves excited in Saturn's rings by the gravitational forcing due to global oscillation modes inside Saturn, \citet{Fuller2014} showed that these were compatible with an interior model that contains {an extended stably stratified region outside a solid core,} that supports gravity modes. In addition, the {ongoing} {Juno} mission {opens the path} to improve our understanding of Jupiter's interior {{\citep[][]{MiguelEtal2016,BoltonEtal2017}}. For instance, it has been estimated from Juno's gravitational measurements that Jupiter's zonal flows extend down to only 3,000 km below cloud level, which constrains the internal rotation of Jupiter \citep{KaspiEtal2018Nature,GuillotEtal2018Nature}. In addition, and more relevant to the present study, it has been suggested by \citet{WahlEtal2017} that the deep} interior structure of Jupiter is consistent with models {in} which the heavy elements of its core are diluted in its envelope. This observation, if confirmed, could corroborate a number of theoretical studies that suggested more complex models of giant planet interiors containing stabilising compositional gradients that hamper large-scale convection in their deep interiors {\citep{Stevenson1982,Stevenson1985,LeconteChabrier2012,VazanEtal2016,GuillotEtal2018}}. This picture significantly departs from the standard three-layer model in which a molecular H/He envelope surrounds a metallic H/He envelope, on top of a rocky/icy core composed of heavy elements \citep[see][]{GuillotEtal2004}. Among the relevant works, some have suggested that regions exhibiting a stable compositional gradient could exist, either at the core boundary due to its erosion \citep{GuillotEtal2004,MazevetEtal2015}, or at the interface between metallic and molecular H/He due to gravitational settling of He droplets in the molecular region \citep{StevensonSalpeter1977,NettelmannEtal2015}. {In particular, the most recent study by \cite{GuillotEtal2018} estimates that compositional gradients could persist from the primordial evolution of Jupiter for $\sim 40\%$ of its mass.}

The presence of a stabilising compositional gradient alongside the destabilising entropy gradient (driving the convection) could trigger oscillatory double-diffusive convection in the form of layered semi-convection \citep[see][for a recent review]{Garaud2018}, in which a large number of well-mixed convective layers are separated by thin stably stratified interfaces \citep[][and references therein]{LeconteChabrier2012, WoodEtal2013}. The associated density profile is nearly constant in the convective steps and undergoes a {sharp} jump in stably stratified interfaces, giving a density staircase-like structure. Such structures are also observed on Earth, for instance in the Arctic Ocean \citep{GhaemsaidiEtal2016} and in geothermally active lakes \citep{WuestEtal2012} ({due to fingering convection}). The number of layers, their thicknesses and the long-term evolution of the staircase are not well constrained based on our current understanding of the physics of these layers and of giant planet interiors more generally. A region of layered semi-convection could be important in the thermal evolution of giant planets, and may explain Saturn's luminosity excess \citep{LeconteChabrier2013} and contribute to the inflated radii of some hot Jupiters \citep[e.g.][]{ChabrierBaraffe2007}. {Finally, we note that \citet{GuillotEtal2018} have estimated the region in which layered semi-convection could potentially be present in Jupiter today as $\sim 10\%$ of its mass.}

{In} this work, {we} study for the first time the impact of layered semi-convection upon the efficiency of tidal dissipation within an idealised Cartesian model. {T}wo tidal components are usually distinguished: the {equilibrium tide}{,} a large-scale flow induced by the quasi-hydrostatic adjustment to the gravitational potential of the perturber \citep[such as the moons of giant planets; see][]{Zahn1966,RemusMathisZahn2012}, {and} the {dynamical tide}, composed of internal waves excited by the perturber \citep{Zahn1975,OgilvieLin2004}. Their restoring forces are buoyancy and the Coriolis acceleration, thus they are often called gravito-inertial waves. Their dissipation by viscosity and thermal diffusion will lead to the long-term rotational, orbital and thermal evolution of the system \citep[e.g.][]{Ogilvie2014}. {In our Cartesian model,} we adopt a tidal-like forcing that is designed to mimic certain aspects of the periodic {forcing of tidal gravito-inertial waves by the} gravitational potential of a moon orbiting a giant planet. The orbital frequency of a moon is (in general) small compared to the dynamical frequency of the planet, and we mostly expect waves in the sub-inertial frequency range (that is, with a frequency less than the Coriolis frequency $2\Omega$, where $\Omega$ is the mean rotation rate of the planet) to be excited resonantly by tidal forcing \citep[e.g.][]{Ogilvie2014}.

Recently, several papers have begun to study how a region of layered semi-convection could modify the propagation of internal (and inertial) waves, and analysed the properties of their associated oscillation modes. \cite{BelyaevQuataertFuller2015} derived the dispersion relation for the free modes of a staircase, showed that regions of layered semi-convection could sustain g-modes, and considered the effects of rotation at the pole and at the equator. \cite{Sutherland2016} studied the transmission of an incident internal wave upon a density staircase embedded in a stably stratified medium under the traditional approximation, which consists of neglecting the horizontal component of the rotation vector in the Coriolis acceleration. Finally, \cite{ABM2017}, hereafter referred to as Paper I, have generalised both of these previous studies by analysing the effects of rotation including the full Coriolis acceleration at any latitude, and studied its effects on the free modes of a density staircase and on the transmission of incident waves. {We found that waves incident on a region of layered semi-convection are preferentially transmitted if their frequencies match those of a free mode of the staircase.}

The present paper focuses on how {the dissipation of} the dynamical tide is affected by the presence of a region of layered semi-convection. {In particular, our underlying motivation is to determine whether the dissipation of tidal waves in a region of layered semi-convection could be significantly enhanced compared to a fully convective (adiabatic) medium. If so, layered semi-convection could play a key role in explaining the high tidal dissipation rates observed in Jupiter \citep{LaineyEtal2009} and Saturn \citep{LaineyEtal2017}, alongside other physical mechanisms such as turbulent friction applied to tidal inertial waves in convective envelopes \citep[e.g.][]{OgilvieLin2004,MathisEtal2016}, the visco-elastic dissipation in rocky/icy dense central regions \citep[e.g][]{RemusMathisZahn2012,GuenelMathisRemus2014}, or the resonant locking of tidal gravito-inertial modes \citep{FullerEtal2016}.}

In Section \ref{sec:NumericalSetUp}, we {introduce} the relevant mathematical and physical aspects of the forcing and dissipation of linear internal (and inertial) waves. Section \ref{sec:layeredCase} presents our study of the layered case. In particular, we explore how the rates of tidal dissipation depend on the properties of the staircase and other control parameters of our model. We compare our results with the fully convective case, since this is the commonly-adopted model of giant planet interiors. Finally, we discuss in Section \ref{sec:Discussion} some of the implications of our results, {particularly for Solar System giant planets}, and we also present our conclusions and discuss some possible directions for future work.

\section{Numerical calculation of the forcing and dissipation of internal waves in a region of layered semi-convection}\label{sec:NumericalSetUp}
%
\subsection{Main assumptions}\label{subsec:MainAssumptions}

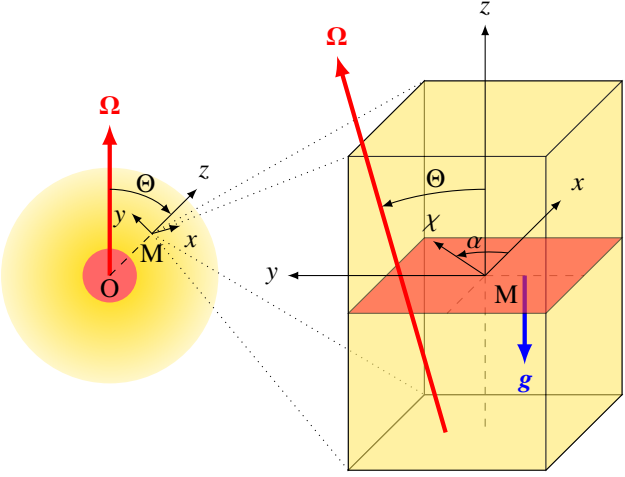
\begin{figure}
\centering
\begin{tikzpicture}[scale=0.52]
\pgfmathsetmacro{\cubex}{5}
\pgfmathsetmacro{\cubey}{8}
\pgfmathsetmacro{\cubez}{5}
\pgfmathsetmacro{\centerx}{-2.4*\cubex}
\pgfmathsetmacro{\centery}{-\cubey/2}
\pgfmathsetmacro{\centerz}{-\cubez/2}
\pgfmathsetmacro{\planetradius}{1.1*\cubex/2}
%
\shadedraw[shading=radial,outer color=fluorescentorange!20,inner color=fluorescentorange,draw=none] (\centerx,\centery,\centerz) circle (\planetradius);
\fill[color=red!60] (\centerx,\centery,\centerz) circle (\planetradius/4);
\node at (\centerx,1.07*\centery-0.05,\centerz) {O};
\draw[load,color=red] (\centerx,\centery,\centerz) -- ++ (0,1.4*\planetradius,0) node[above]{$\bm{\Omega}$};
\draw[dotted] (\centerx+0.55*\planetradius/1.4142,\centery+0.55*\planetradius/1.4142,\centerz) -- (-\cubex,0,-\cubez);
\draw[dotted] (\centerx+0.55*\planetradius/1.4142,\centery+0.55*\planetradius/1.4142,\centerz) -- (-\cubex,0,0);
\draw[dotted] (\centerx+0.55*\planetradius/1.4142,\centery+0.55*\planetradius/1.4142,\centerz) -- (-\cubex,-\cubey,-\cubez);
\draw[dotted] (\centerx+0.55*\planetradius/1.4142,\centery+0.55*\planetradius/1.4142,\centerz) -- (-\cubex,-\cubey,0);
\draw[dashed] (\centerx,\centery,\centerz) -- ++ (0.55*\planetradius/1.4142,0.55*\planetradius/1.4142,0);
\draw[axis] (\centerx+0.55*\planetradius/1.4142,\centery+0.55*\planetradius/1.4142,\centerz) -- ++ (\planetradius/2.4,\planetradius/2.4,0) node[above]{$~~z$};
\draw[axis] (\centerx+0.55*\planetradius/1.4142,\centery+0.55*\planetradius/1.4142,\centerz) -- ++ (-\planetradius/5.3,\planetradius/5.3,0);
\node at (\centerx+0.25,\centery+1.45,\centerz) {$y$};
\draw[axis] (\centerx+0.55*\planetradius/1.4142,\centery+0.55*\planetradius/1.4142,\centerz) -- ++ (\planetradius/5.3,0,-\planetradius/5.3) node[below]{$~~~x$};
\node at (\centerx+0.55*\planetradius/1.4142,\centery+0.55*\planetradius/1.4142-0.5,\centerz) {M};
\draw[axis,color=black] (\centerx,\centery+0.8*\planetradius,\centerz) arc (90:45:0.8*\planetradius);
\node at (\centerx+\planetradius/3,\centery+0.855*\planetradius,\centerz) {$\Theta$};
%
\draw[dashed] (-\cubex/2,-\cubey/2,-\cubez/2) -- ++ (\cubex/2,0,0);
\draw[dashed] (-\cubex/2,-\cubey/2,-\cubez/2) -- ++ (0,0,\cubez/2);
\draw[dashed] (-\cubex/2,-\cubey/2,-\cubez/2) -- ++ (0,-\cubey/2,0);
\draw[black] (0,0,-\cubez) -- ++(-\cubex,0,0) -- ++(0,-\cubey,0) -- ++(\cubex,0,0) -- cycle;
\draw[black] (-\cubex,0,0) -- ++(0,0,-\cubez) -- ++(0,-\cubey,0) -- ++(0,0,\cubez) -- cycle;
\draw[black] (0,-\cubey,0) -- ++(-\cubex,0,0) -- ++(0,0,-\cubez) -- ++(\cubex,0,0) -- cycle;
\draw[black,fill=fluorescentorange!60,opacity=0.6] (0,0,0) -- ++(-\cubex,0,0) -- ++(0,-\cubey,0) -- ++(\cubex,0,0) -- cycle;
\draw[black,fill=fluorescentorange!60,opacity=0.6] (0,0,0) -- ++(0,0,-\cubez) -- ++(0,-\cubey,0) -- ++(0,0,\cubez) -- cycle;
\draw[black,fill=fluorescentorange!60,opacity=0.6] (0,0,0) -- ++(-\cubex,0,0) -- ++(0,0,-\cubez) -- ++(\cubex,0,0) -- cycle;
\draw[load,color=blue] (-0.9*\cubex/3,-\cubey/2,-\cubez/2) -- ++ (0,-0.85*\cubey/3,0) node[below]{$\bm{g}$};
\draw[black,fill=red!80,opacity=0.6] (0,-\cubey/2,0) -- ++(-\cubex,0,0) -- ++(0,0,-\cubez) -- ++(\cubex,0,0) -- cycle;
\draw[load,color=red] (-3.5*\cubex/5,-\cubey,-\cubez/2) -- ++ (-0.55*\cubex,1.2*\cubey,0) node[above]{$\bm{\Omega}$};
\draw[axis] (-\cubex/2,-\cubey/2,-\cubez/2) -- ++ (-1.25*0.7071*\cubex/2,0,-1.3*0.7071*\cubez/2) node[above]{$\chi$};
\draw[axis,color=black] (-\cubex/2,-\cubey/2,-4*\cubez/5) arc (84:103.18:4);
\node at (-2.12*\cubex/3,-0.98*\cubey/2,-4.5*\cubez/5) {$\alpha$};
\draw[axis] (-\cubex/2,-\cubey/2,-\cubez/2) -- ++ (-\cubex,0,0) node[left] {$y$};
\draw[axis] (-\cubex/2,-\cubey/2,-\cubez/2) -- ++ (0,0,-\cubez) node[above right] {$x$};
\draw[axis] (-\cubex/2,-\cubey/2,-\cubez/2) node[below right]{M} -- ++ (0,0.8*\cubey,0) node[above] {$z$};
\draw[axis,color=black] (-\cubex/2,-0.9*\cubey/4,-\cubez/2) arc (90:109.15:\cubey);
\node at (-3*\cubex/4,-0.95*\cubey/5,-\cubez/2) {$\Theta$};
\end{tikzpicture}
\caption[Local Cartesian box]{{Left:} global view of a giant planet: the gaseous envelope (in yellow, the shading denoting density), lies on top of {the core} (in red). {Right:} magnified picture of the local Cartesian box, centered on a point M of a giant planet envelope, corresponding to a colatitude $\Theta$. The box is tilted with respect to the spin axis, and its vertical axis is $z$, corresponding to the local radial direction, is thus anti-aligned with gravity. The $x$ and $y$ axes correspond to the local azimuthal and latitudinal directions, respectively, while the $\chi$ axis makes an angle $\alpha$ with respect to the $x-$axis.}
\label{fig:cartesianBox}
\end{figure}

Our main assumptions are the same as in Paper I, except that we now take into account dissipative processes. We adopt the Boussinesq approximation in a local Cartesian model {\citep{GerkemaShrira2005,MathisNeinerTranminh2014}} that represents a small-patch of a giant planet. We centre our box on a point M of the gaseous envelope (see Fig. \ref{fig:cartesianBox}). {We thus neglect the sphericity of the problem as a first step {\citep[see e.g. the appendix of][for the cases of pure inertial waves, gravito-inertial waves, and gravito-inertial waves in the layered case, respectively]{OgilvieLin2004,AuclairDesrotourEtal2015,ABM2017}}}. {Indeed,} we wish to study the dissipation of short-wavelength internal waves by viscosity and thermal diffusion, which are represented here by a constant kinematic viscosity $\nu$, and a constant thermal diffusivity $\kappa$. {In the case of convective layers, these coefficients represent an effective viscosity/diffusivity that accounts for turbulent friction acting on tidal waves \citep[e.g][]{Zahn1966,Zahn1989,GoldreichKeeley1977a,OgilvieLesur2012,MathisEtal2016}.}

We adopt the same notation as Paper I, in particular our local system of coordinates $(x,y,z)$ corresponds to the local azimuthal, latitudinal and radial directions, respectively. The rotation vector $\bm{\Omega}$ is inclined by an angle $\Theta$ with respect to the gravity vector $\bm{g}$, which is (anti-)aligned with the vertical direction. Thus, the latitudinal and vertical components of the rotation vector are, respectively, 
\begin{align}
\tilde{f} &= 2\Omega\sin\Theta,\label{eq:ftilde}\\
f &= 2\Omega\cos\Theta,\label{eq:f}
\end{align}
so that $2\bm{\Omega} = (0, \tilde{f}, f).$ {We follow \cite{GerkemaShrira2005} and we introduce a reduced horizontal coordinate,} $\chi$, which makes an angle $\alpha$ with respect to the $x-$axis: 
\begin{equation}
  \chi =x\cos\alpha + y\sin\alpha.
\end{equation}
{This will allow us to treat the problem within a two-dimensional framework.} {We finally define
\begin{equation}
\tilde{f}_{\text{s}} = \tilde{f}\sin\alpha.
\end{equation}
}

\subsection{Equations of motion and energetics}\label{sec:equationsofmotion}
We study the linear excitation of gravito-inertial waves subject to dissipative processes, namely viscosity and thermal diffusion. We now include an external body forcing $\bm{F}$, with components $(F_x,F_y,F_z)$ in the local Cartesian model. {We note that tidal gravito-inertial waves are not forced directly by the tidal potential, but by the Coriolis acceleration applied to the equilibrium tide \citep[see e.g.][and references therein]{Ogilvie2014}.} We focus on the linear tidal response in this study, neglecting the effects of fluid nonlinearities on the dissipation (or excitation) of waves \citep{JouveOgilvie2014,FavierEtal2014}. This is likely to be an appropriate assumption for studying the excitation of waves in giant planets excited by their natural satellites, though it is possible that nonlinear effects could still play some role in damping short-wavelength waves, which is neglected here.

The linearised components of the momentum equation in the Boussinesq approximation are
\begin{align}
D_{\nu} u - fv + \tilde{f}w &= -\frac{1}{\rho_0} \dfrac{\partial p}{\partial x} + F_x \label{momx}\\
D_{\nu} v + fu &= -\frac{1}{\rho_0} \dfrac{\partial p}{\partial y} + F_y\label{momy}\\
D_{\nu} w - \tilde{f}u &= -\frac{1}{\rho_0} \dfrac{\partial p}{\partial z} + b + F_z,\label{momz}
\end{align}
where $u$, $v$ and $w$ are the components of the velocity perturbation in the local azimuthal, latitudinal and radial directions, respectively, $\rho_0$ is a constant reference value for the density, $p$ is the pressure fluctuation, and 
\begin{equation}
D_{\nu} = \partial_t - \nu \nabla^2.
\end{equation}
The continuity equation is
\begin{equation}
\dfrac{\partial u}{\partial x} + \dfrac{\partial v}{\partial y} + \dfrac{\partial w}{\partial z} = 0. \label{cont}
\end{equation}
Finally, the {thermal energy equation} is
\begin{equation}
D_{\kappa} b + N^2 w = 0 \label{ener},
\end{equation}
where
\begin{equation}
D_{\kappa} = \partial_t - \kappa \nabla^2,
\end{equation}
and
\begin{equation}
b = -g\frac{\rho}{\rho_0}
\end{equation}
is the fluid buoyancy, $\rho$ is the density fluctuation, and $N^2(z)$ is the squared buoyancy frequency, whose $z$-dependence is chosen to model a layered density structure (see Section \ref{sec:model}). 

From the above set of equations, we derive in Appendix \ref{app:PoincareEq} the {forced Poincar\'e equation},
\begin{equation}
D_{\kappa}D_{\nu}^2 \nabla^2 w + D_{\kappa}(\bm{f} \cdot \bm{\nabla})^2 w + D_{\nu}\left[N^2 \nabla_{\perp}^2\right] w
= \bm{\mathcal{O}} \cdot (\bm{\nabla} \times \bm{F}),
\label{eq:forcedPoincare Eq.DOC}
\end{equation}
where the operator $\bm{\mathcal{O}}$ is
\begin{equation}
\bm{\mathcal{O}} \equiv D_{\kappa}
\left(
      D_{\nu}\dfrac{\uppartial}{\uppartial y},
      -D_{\nu}\dfrac{\uppartial}{\uppartial x},
      -\bm{f}\cdot\bm{\nabla}
\right).
\label{eq:operator}
\end{equation}
Equation (\ref{eq:forcedPoincare Eq.DOC}) governs the spatio-temporal evolution of the vertical velocity of gravito-inertial waves driven by a prescribed body force $\bm{F}$, in the presence of dissipative mechanisms. Our adopted body force will be designed to mimic certain aspects of tidal forcing.

We wish to understand how efficiently tidally-forced waves are dissipated in a region of layered semi-convection. From equations (\ref{momx})--(\ref{ener}), we thus derive an energy balance equation
\begin{equation}
\frac{\text{d}\bar{E}}{\text{d}t} = -\frac{1}{V}\int_S \bm{\Pi}\cdot\text{d}\bm{S} + \bar{D}_{\text{visc}} + \bar{D}_{\text{ther}} + \bar{I},
\label{eq:energy_balance}
\end{equation}
where $\bar{E}$ is the total (pseudo-)energy of the wave, the sum of kinetic and (available) potential energies (both spatially-averaged over the box):
\begin{equation}
E_{\text{k}} = \frac{1}{V} \int_{\mathcal{V}} \frac{1}{2}\rho_0|\bm{u}|^2\,\text{d}\mathcal{V},
\label{eq:def_kineticEnergy}
\end{equation}
and
\begin{equation}
E_{\text{p}} =
\left\{
\begin{array}{ccc}
\displaystyle \frac{1}{V} \int_{\mathcal{V}} \frac{1}{2}\rho_0\frac{b^2}{N^2} ~ & \text{if} & ~ N^2 \ne 0,\\[2.5mm]
\displaystyle 0 ~ & \text{if} & ~ N^2=0,
\end{array}
\right.
\label{eq:def_potentialEnergy}
\end{equation}
respectively. We also define $\bm{\Pi} = p\bm{u}$, the flux density (flux per unit area) of energy, and the volume-averaged dissipation by viscosity and thermal diffusion:
\begin{align}
\bar{D}_{\text{visc}} &= \frac{1}{V}\int_{\mathcal{V}} \rho_0\left( \nu \bm{u} \cdot \nabla^2 \bm{u} \right)\,\text{d}\mathcal{V},\\
\bar{D}_{\text{ther}} &= \left\{
\begin{array}{ccc}
\displaystyle \frac{1}{V}\int_{\mathcal{V}} \rho_0 \left(\frac{\kappa}{N^2}b \nabla^2b\right)\,\text{d}\mathcal{V} ~ & \text{if} & ~ N^2 \ne 0,\\[2.5mm]
0 ~ & \text{if} & ~ N^2 = 0,
\end{array}
\right.
\end{align}
respectively. Finally, the mean rate of energy injection by the forcing is
\begin{equation}
\bar{I} = \frac{1}{V}\int_{\mathcal{V}} \rho_0(\bm{u}\cdot\bm{F})\,\text{d}\mathcal{V}.
\end{equation}

Our goal is to calculate numerically the volume-averaged rates of viscous and thermal dissipation, $\bar{D}_{\text{visc}}$ and $\bar{D}_{\text{ther}}$, respectively, and the averaged total dissipation rate,
\begin{equation}
\bar{D} = \bar{D}_{\text{visc}} + \bar{D}_{\text{ther}},
\label{eq:Dtot}
\end{equation}
as various parameters of our problem are varied. In particular, we will compute the frequency dependence of the dissipation by varying the frequency of the forcing, $\omega$, and thus obtain what we will refer to as ``dissipation spectra". We are also interested in computing frequency-averaged dissipation, following \cite{Ogilvie2013},
\begin{equation}
\left< \bar{D}\right> = \int_{-\infty}^{+\infty}\bar{D}(\omega) \,\frac{\text{d}\omega}{\omega},
\label{eq:meanD}
\end{equation}
which provides a measure of the dissipation at low frequencies. This quantity will be useful for studying how the dissipation varies with the parameters of our problem, including the properties of the background density staircase (see Section \ref{sec:model}). The fact that this is weighted with the inverse of the tidal frequency, naturally makes it a frequency-averaged measure of the dissipation in the low-frequency range corresponding to inertial waves, which is usually the relevant range for tidal forcing \citep[e.g.][]{Ogilvie2014}.
{Moreover, the final result will strongly depend on the dependence of the forcing to the tidal frequency.}
{We also note that the expression given by Eq. (\ref{eq:meanD}) has been used for applications to giant planets \citep{GuenelMathisRemus2014} and stars \citep{Mathis2015,GalletEtal2017,BolmontEtal2017}, and  provides a representative order of magnitude of the tidal dissipation, so that its consequences on the evolution of planetary systems can be studied \citep{BolmontMathis2016,DamianiMathis2018}.}

\subsection{Numerical statement of the problem}
We consider our variables to vary as
\begin{equation}
a(x,y,z,t) = \Re\left\{{A}(z) \exp\left[\text{i} (k_x x + k_y y - \omega t)\right]\right\},
\label{eq:transformationForced}
\end{equation}
where the wavenumbers in the horizontal direction, $k_x$ and $k_y$, and the frequency $\omega$, are the same as that of the tidal-like forcing {$\bm{F}$. The latter is taken} to be $2\uppi$
periodic in the $x$ and $y$ directions, and in time. {Thus, we write}
\begin{equation}
\bm{F} = \tilde{\bm{F}}(z) \exp \left[ \text{i} (k_{\perp}\chi - \omega t) \right].
\end{equation}
Here, {the vector $\tilde{\bm{F}} = (\tilde{F}_{x}, \tilde{F}_{y}, \tilde{F}_{z})$ contains} the Fourier components of $\bm{F}$ in the local azimuthal, latitudinal and radial directions, respectively.

We note that the frequency $\omega$ is not necessarily equal to the orbital frequency of a companion. Indeed, in the case of a circular aligned orbit, the dominant component of the tidal potential has $\omega= 2(n-\Omega)$ (where $n$ is the orbital frequency), which is not equal to $n$ in general. In addition, it is appropriate for us to solve only for the $k_x$, $k_y$ and $\omega$ of the forcing because we are considering a linear problem (so that all horizontal wavenumbers and frequencies are uncoupled). In the case of an eccentric or inclined orbit, several frequencies should instead be considered (e.g.~Zahn 1966a, 1977; Mathis \& Le Poncin-Lafitte 2009; Ogilvie 2014).

\subsubsection{Vertically periodic boundary conditions}
\label{vertperiodic}
{For the first set of calculations, we have assumed quantities to be periodic in the vertical direction}, which is equivalent to considering part of a more vertically-extended staircase. To solve the system of equations (\ref{momx})--(\ref{ener}) numerically, we can therefore use a Fourier collocation method \citep{Boyd2000}. This method assumes that approximate solutions are represented as a discrete Fourier series that matches the exact solution on a set of collocation points, defined by
\begin{equation}
z_n = z_{\text{i}} + (z_{\text{o}}-z_{\text{i}})\frac{n}{N_z} ~~ \text{for} ~~ n=\{0,\dots,N_z-1\},
\end{equation}
where $N_z$ is the number of grid points, and $z_{\text{i}}$ and $z_{\text{o}}$ define the inner and outer edge of the box, respectively. Here we choose $z_i=-L_z/2$ and $z_o=L_z/2$. A spectral collocation method is used in preference to finite differences to approximate the derivatives because spectral methods are more accurate for smooth solutions, allowing us to use fewer grid points to obtain the same accuracy, which is computationally more efficient.
Vertical derivatives are performed using the derivative matrix $\bm{\mathsf{D}}_1$ \citep[see e.g. Appendix F of][]{Boyd2000}, defined by
\begin{equation}
(\bm{\mathsf{D}}_1)_{ij} = \left\{
\begin{array}{lcc}
\dfrac{1}{2} (-1)^{i+j} \cot\left(\dfrac{z_i-z_j}{2}\right) & ~~ \text{if} ~~ & i \neq j,\\[5mm]
0 & ~~ \text{if} ~~ & i = j.
\end{array}
\right.
\label{eq:derivativeMatrix}
\end{equation}
We adopt a collocation method, rather than a Fourier Galerkin method because this allows us to solve for modes in a spatially-varying (with $z$) density structure more efficiently.

While considering periodic boundary conditions in the vertical is relevant for studying {global modes propagating in} a portion of a more vertically-extended staircase, this assumption does exclude certain effects. In particular, this model prevents reflection of internal waves from a solid core, which can be important in the geometrical focusing of internal wave beams along wave attractors \cite[e.g.][]{OgilvieLin2004}. This should be considered in an equivalent study in a global geometry. Nevertheless, here we extend our study in this Cartesian model by modifying boundary conditions.\\

\subsubsection{Vertically rigid and stress-free boundary conditions}
{Indeed, to address this point,} we also consider a model with the same setup as above, {but with impenetrable, stress-free boundary conditions in the vertical. For clarity, we will refer to them as \textit{rigid boundary conditions} hereafter.} This model permits the reflection of waves from the boundaries, and leads to a modification of the global modes{, as in the case of periodic boundary conditions}. We implement these conditions by using a Chebyshev collocation method instead of a Fourier method in the vertical direction. Our setup assumes impenetrable, stress-free boundary conditions, with zero buoyancy perturbation at the upper and lower boundaries. Namely, at $z=z_i$ and $z=z_o$, we impose that:
\begin{equation}
\left\{\begin{array}{rcl}
\partial_z u & = & 0,\\
\partial_z v & = & 0,\\
w & = & 0,\\
b & = & 0.
\end{array}\right.
\end{equation}
We otherwise solve the same system of equations as in \S~\ref{vertperiodic} at the Gauss-Lobatto points, such that
\begin{equation}
z_n=\frac{1}{2}\left[\left(1+x_n\right)z_o+(1-x_n)z_i\right],
\end{equation}
where
\begin{equation}
x_n = \cos\left(\frac{\pi n}{N_z}\right), \;\;\;\;\;\; n = 1,\dots,N_z.
\end{equation}
Derivatives are calculated using a different derivative matrix $\bm{\mathsf{D}}_1$ \citep[see e.g. Appendix F of][]{Boyd2000} than in \S~\ref{vertperiodic}, which is now defined by
\begin{equation}
(\bm{\mathsf{D}}_1)_{ij} = \left\{
\begin{array}{lcc}
(-1)^{i+j} \dfrac{p_i}{p_j(x_i-x_j)} & ~~ \text{if} ~~ & i \neq j,\\[5mm]
-\dfrac{x_j}{2(1-x_j^2)}             & ~~ \text{if} ~~ & i = j\,\,; ~~ 0 < j < N,\\[5mm]
-\dfrac{1}{6}(1+2N^2)                & ~~ \text{if} ~~ & i = j = N,\\[5mm]
\dfrac{1}{6}(1+2N^2)                & ~~ \text{if} ~~ & i = j = 0,
\end{array}
\right.
\label{eq:derivativeMatrixChebyshev}
\end{equation}
where $p_0=p_N=2$, and $p_j=1$ otherwise.

This setup represents a plane-parallel model of an extended giant planet envelope filled with a region of layered semi-convection. The reflection of internal waves from the core and surface are then allowed. This model may be more realistic than the one considered in \S~\ref{vertperiodic}, but for simplicity we continue to neglect global curvature effects that would be present in spherical geometry, and continue to adopt the Boussinesq approximation, which prevents us from studying realistic planetary density profiles (where the density should vary over several orders of magnitude).

\subsection{Modeling the layered structure and the forcing}\label{sec:model}
The assumption of periodicity in the vertical direction is appropriate if we consider our model to represent part of a more vertically-extended density staircase. We {can then} consider our staircase to have a periodicity of $m$ steps. We also consider rigid upper and lower boundaries, which is appropriate if we consider our model to represent a plane-parallel layer of a giant planet entirely filled with layered semi-convective steps and containing $m$ stably-stratified interfaces.

\subsubsection{Layered structure and buoyancy frequency profile}
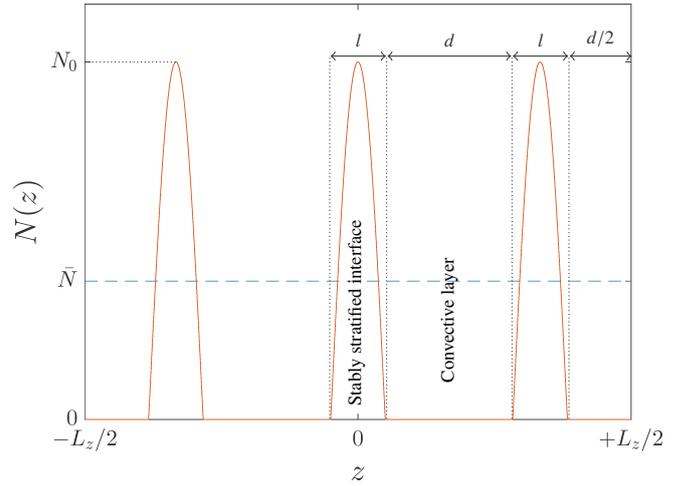
\begin{figure}
\centering
\begin{tikzpicture}[scale=1.11815]
\pgfmathsetmacro{\lag}{0.01}
\node[anchor=south west,inner sep=0] at (0,0)
    {\includegraphics[width=0.95\linewidth]{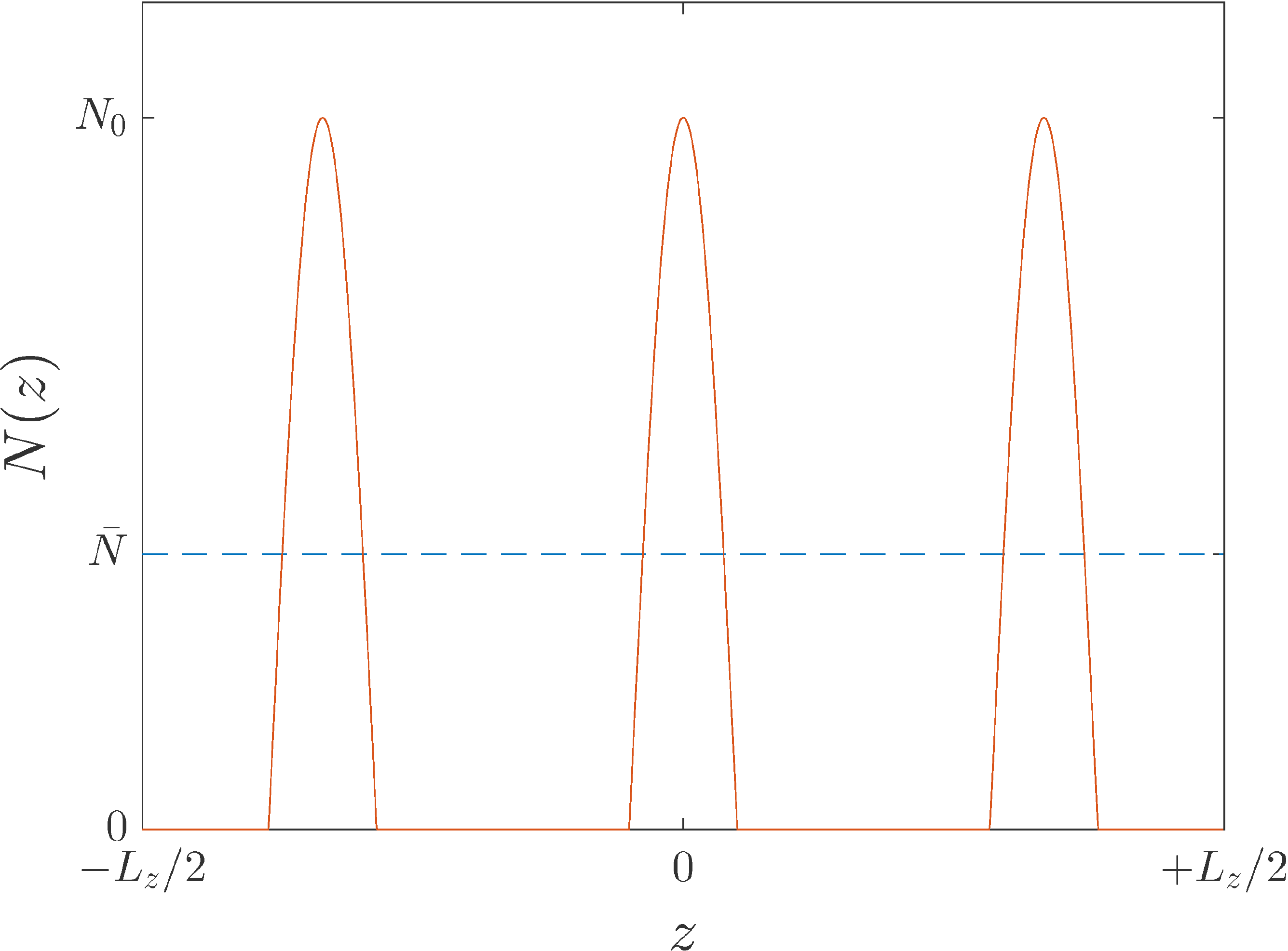}};
\draw[densely dotted,color=black!80] (0.8475,4.23+0.73) -- ++ (1.08,0);
\draw[densely dotted,color=black!80] (3.73,0.73) -- ++ (0,4.3);
\node[rotate=90] at (4.05,1.99) {\scriptsize Stably stratified interface};
\draw[densely dotted,color=black!80] (4.4,0.73) -- ++ (0,4.3);
\node[rotate=90] at (5.14,1.9) {\scriptsize Convective layer};
\draw[densely dotted,color=black!80] (5.875,0.73) -- ++ (0,4.3);
\draw[densely dotted,color=black!80] (5.875+4.4-3.73,0.73) -- ++ (0,4.3);
\draw[<->,color=black!80] (3.73+\lag,4.3+0.73) -- node[above]{\scriptsize $l$} (4.4-\lag,4.3+0.73);
\draw[<->,color=black!80] (4.4+\lag,4.3+0.73) -- node[above]{\scriptsize $d$} (5.875-\lag,4.3+0.73);
\draw[<->,color=black!80] (5.875+\lag,4.3+0.73) -- node[above]{\scriptsize $l$} (5.875+4.4-3.73-\lag,4.3+0.73);
\draw[<->,black!80] (5.875+4.4-3.73+\lag,4.3+0.73) -- node[above]{\scriptsize $d/2$} (5.875+4.4-3.73+0.7475-\lag,4.3+0.73);
\end{tikzpicture}\\[-3pt]
\caption[Numerical set up of the buoyancy frequency profile]{Buoyancy frequency $N(z)$ for an example with 3 steps. Convective layers of size $d$, within which $N(z)=0$, are separated by stably stratified interfaces of size $l$, within which $N(z)>0$, given by equation (\ref{eq:Nprofile}). The amplitude of the peaks, $N_0$, is calculated so that the mean stratification equals $\bar{N}$. Here, $N_0\approx 2.58 \bar{N}$.}
\label{fig:Nprofile}
\end{figure}

In a layered structure, the buoyancy frequency $N$ (and accordingly the background density gradient) is not uniform and can vary with $z$ on rather short length scales. Unlike in \citetalias{ABM2017}, we adopt a smooth buoyancy frequency profile, which is advantageous numerically. We take a profile like that displayed on Fig. \ref{fig:Nprofile}, to model the alternation of convective and stably stratified layers.
Namely, we take
\begin{equation}
N^2(z)=
\left\{
\begin{array}{cl}
\dfrac{N_0^2}{2}\left(1+\cos\left(2\uppi\dfrac{(z-z_j)}{l}\right)\right) & ~ \text{if} ~~~~ |z-z_j|<l/2\\[4mm]
0 & ~ \text{otherwise},
\end{array}
\right.
\label{eq:Nprofile}
\end{equation}
so that stably-stratified interfaces correspond to positive values of the squared buoyancy frequency, while this is taken to be zero in convective layers (so that they are isentropic and well mixed). We define $m$ to be the number of steps in the domain (or equivalently the number of interfaces), $l$ to be the size of the stably stratified layers, and $z_j$ to be the position of the $j$-th interface, defined by 
\begin{equation}
z_j = -\frac{(m+1)}{2m}L_z + \frac{(j-1)}{m}L_z.
\end{equation}
The size of the convective layers is then $d = (L_z - ml)/m$. This placement of the interfaces ensures that the distance between adjacent interfaces equals $d$, except at both ends of the domain in $z$. We also define the aspect ratio $\varepsilon$, as the ratio between the size of the stably stratified interfaces, $l$, and the size of the convective layers, $d$,
\begin{equation}
\varepsilon = \frac{l}{d}.
\label{eq:eps}
\end{equation}
This parameter is expected to be small in the planetary regime. The step size, $d$, is expected to be smaller than the density (or pressure) scaleheight, $H$, with a ratio $d/H$ that is likely to be in the range $10^{-6} \lesssim d/H \lesssim 1$ \citep{LeconteChabrier2012,NettelmannEtal2015}.

The amplitude of the buoyancy frequency, $N_0$, is calculated to obtain the prescribed mean stratification 
\begin{equation}
\bar{N}^2 \equiv \int_{-z_{\text{i}}/L_z}^{+z_{\text{o}}/L_z} N^2(\zeta)\,\text{d}\zeta,
\label{eq:Nmean}
\end{equation}
where the dimensionless variable $\zeta = z/L_z$. This gives
\begin{equation}
N_0 = \bar{N} \left(\frac{m}{2}\frac{l}{L_z}\right)^{-1/2} = \sqrt{2}\bar{N} \left(\frac{1+\varepsilon}{\varepsilon}\right)^{1/2} .
\label{eq:amplitudeBVfreq}
\end{equation}
Figure \ref{fig:Nprofile} shows an example profile of $N(z)$ with three steps, for which $N_0 \approx 2.58\bar{N}$. 

\subsubsection{Box parameters, and dimensionless numbers}
The vertical extent of the box is chosen such that $L_z=1$, with the domain extending from $z_{\text{i}}=-1/2$ to $z_{\text{o}}=1/2$. The rotation rate is such that $2\Omega = 1$, and the spin axis is chosen here to make an angle $\Theta = \uppi/4$ with respect to the direction of gravity, in order to study the mid-latitude. By choosing these parameters, we have defined our units of length to be $L_z$ and time to be $(2\Omega)^{-1}$. We finally set $\rho_0=1$ to define our unit of mass. 

To quantify the relative importance of diffusive processes, we use the Ekman number,
\begin{equation}
\text{E} = \frac{\nu}{2\Omega L_z^2},
\label{eq:Ekman}
\end{equation}
and its equivalent for thermal dissipation,
\begin{equation}
K = \frac{\kappa}{2\Omega L_z^2}.
\label{eq:EkmanThermal}
\end{equation}
Unless specified otherwise, we set $\text{E}=K$, thus giving a Prandtl number $\text{Pr}\equiv\nu/\kappa=\text{E}/K=1$. In planetary interiors, we expect smaller values for Pr, {which can typically be of order $10^{-2}$ or smaller \citep{WoodEtal2013}. Similarly, $\text{E}$ and $K$ are expected to reach much smaller values in reality than we have chosen here. Typically, the microscopic viscosity estimated by the models of \citet{GuillotEtal2004} correspond to E of the order $10^{-18}$ \citep{OgilvieLin2004}. However,} such values are computationally inaccessible, with smaller values making the problem much more computationally demanding. Our intention is to probe the physics using accessible parameter values, with the hope that we can extrapolate to the astrophysical regime. {On the other hand, in the case of convective layers, a Prandtl number associated with an effective turbulent viscosity and thermal diffusivity may be of order 1.}

In order to quantify the importance of stratification relative to rotation, we also define the dimensionless number
\begin{equation}
\bar{S}_{\Omega} = \frac{\bar{N}}{2\Omega}.
\label{eq:Nover2Omega}
\end{equation}
Unless specified otherwise, we take $\bar{N}/2\Omega=10$. The relevant value for this parameter is uncertain. For example, in the Arctic ocean, \citet{GhaemsaidiEtal2016} found that $\bar{N} \sim$ 70 mrad/s, while $2\Omega_{\text{Earth}} \sim$ 0.07 mrad/s (and $2\Omega_{\text{Jupiter}} \sim$ 0.35 mrad/s). However, it is unclear from theory or simulation what this parameter could be in the deep interiors of giant planets. \cite{Fuller2014} adopts a typical value of $\bar{N}/2\Omega_{\text{Saturn}} \sim 5$ to model a stably stratified region outside the core of Saturn. In Sect. \ref{sec:ParameterSpace/Nmean}, we will explore how tidal dissipation varies as a function of $\bar{S}_{\Omega}$.

\subsubsection{Forcing term}\label{sec:forcing}
We adopt a body force vector 
\begin{equation}
\bm{F} = \tilde{F}_y \exp\left[\text{i}(k_{\perp} \chi + k_zz-\omega t)\right] \hat{\bm{e}}_y,
\label{eq:forcing}
\end{equation}
with $\tilde{F}_y = 1$ and $k_{\perp}=k_z=2\uppi$. 
We recall that $\chi = x \cos\alpha + y \sin\alpha$, where we have chosen $\alpha=\uppi/2$. Our choice of $\bm{F}$ is somewhat academic, but is designed to mimic aspects of large-scale tidal forcing, i.e. forcing of waves by the equilibrium tide, which acts as an effective force driving the dynamical tide. In reality, this is radially node-less (so not oscillatory in $z$), but we {first} take it to be periodic in $z$ with the longest wavelength for numerical convenience.
{In order to relax the assumption of periodicity of the forcing with rigid boundaries, we have also performed calculations for which the forcing term was given a linear dependence in $z$, namely $\bm{F} = \tilde{F}_y\, z\, \hat{\bm{e}}_y$. This is a more realistic approximation to the driving of waves by the equilibrium tide in a local model. However, this did not produce any significant modification to our results with a periodic forcing in $z$ (in general). We will therefore focus on cases with periodic forcing in $z$.}

In addition, we have explored the effect of modeling a frequency-dependent expression for the forcing term when computing frequency-averaged dissipation rates (see Section \ref{sec:ParameterSpace}) according to Eq. (\ref{eq:meanD}).
{This is physically more realistic in the sense that the equilibrium tide \citep[e.g][]{Zahn1966,RemusMathisZahn2012}, which is responsible for forcing the tidal waves, has a velocity amplitude that is proportional to the tidal frequency $\omega$ \citep[e.g.][]{OgilvieLin2004,Ogilvie2013}. The resulting forcing of the dynamical tide itself is composed of two terms: the acceleration of the equilibrium tide, and the Coriolis acceleration applied to the equilibrium tide \citep[see e.g. Eq.~(B6) of][]{Ogilvie2005}. The first has a frequency dependence in $\omega^2$, while the second has a frequency dependence in $\omega$ \citep[e.g.][]{OgilvieLin2004,Ogilvie2013}. When computing frequency-averaged dissipation spectra in Section \ref{sec:ParameterSpace}, we choose to take $\tilde{F}_y = \omega$ (instead of 1 for calculations of dissipation spectra) to account for the term that dominates at low (sub-inertial) tidal frequencies, but in principle both terms should be taken into account. This property also ensures that the dissipation goes to 0 when the tidal frequency $\omega$ vanishes, such as when spin-orbit synchronisation occurs.}

We have chosen a simplistic expression for the forcing as a first step to study the problem of tidal dissipation in layered semi-convection. This is because our main goal is to clearly identify the physical effects of a layered density structure on tidal dissipation. Choosing a more general form for $\bm{F}$, while it might be more realistic, would result in more resonant peaks, making the physical interpretation more challenging. We defer calculations adopting such a more realistic tidal forcing to a future study in spherical geometry.

\subsection{A test case in a uniformly stratified medium}\label{sec:uniform}
In order to check the validity of our numerical code using both the Fourier {and Chebyshev} collocation methods, we perform two test cases with uniform buoyancy frequency profiles, as in the Appendix of \cite{OgilvieLin2004} and in \cite{AuclairDesrotourEtal2015}. We have verified that our code accurately agrees with their results in the appropriate cases. For the uniform case {with vertically periodic boundary conditions}, a Fourier transform can also be used in the $z$-direction and the system (\ref{momx})--(\ref{ener}) can be solved analytically.

\begin{figure*}
\vspace{5pt}
\centering
\begin{tikzpicture}[scale=0.495/0.48]
\node[anchor=south west,inner sep=0] at (0,0)
    {\includegraphics[width=0.495\linewidth]{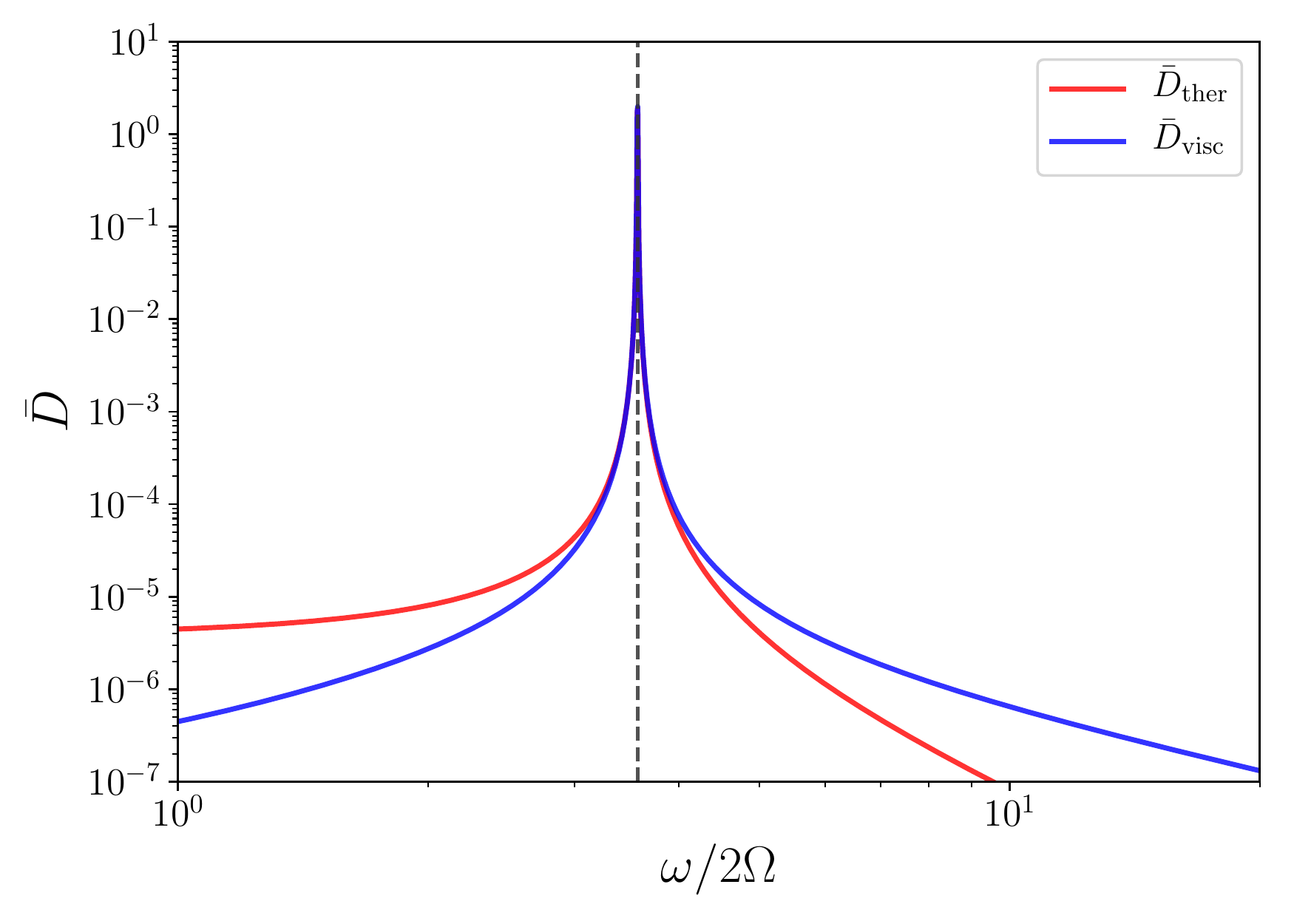}};
\node[rotate=90] at (4.15,1.75) {\small $\omega=\tilde{\omega}^{\text{(GIW)}}$};
\end{tikzpicture}
\begin{tikzpicture}[scale=0.495/0.48]
\node[anchor=south west,inner sep=0] at (0,0)
	{\includegraphics[width=0.495\linewidth]{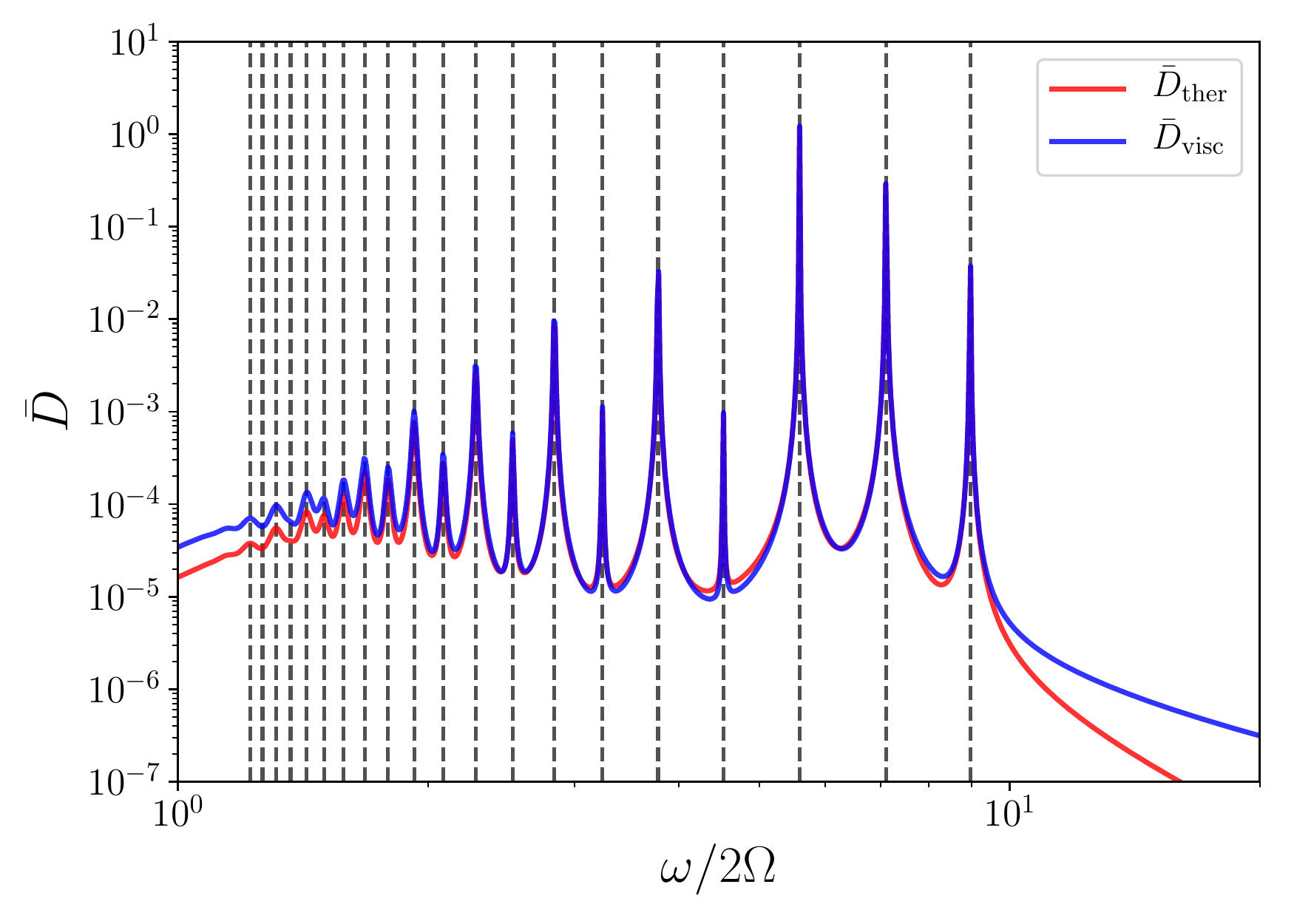}};
\node[rotate=90] at (6.44,5.55) {\small $n=1$};
\node[rotate=90] at (5.85,5.55) {\small $n=2$};
\node[rotate=90] at (5.28,5.55) {\small $n=3$};
\node[rotate=0] at (4.75,5.5) {\normalsize $\dots$};
\end{tikzpicture}\\[-10pt]
\caption{Viscous (in blue) and thermal (in red) dissipation spectra in a uniformly stably stratified medium with $N(z) = \bar{N} = 10\Omega$ throughout the domain {in the case of periodic (\textit{left panel}) and rigid (\textit{right panel}) boundary conditions. \textit{Left panel:} the resonant peak is centered on the single gravito-inertial mode that is resonant with the forcing, with frequency $\tilde{\omega}^{\text{(GIW)}}$ given by Eq. (\ref{eq:om_giw_}). \textit{Right panel:} {rigid boundary conditions do not perfectly excite only a single global mode, they instead excite many modes that match the roots of the dispersion relation of gravito-inertial waves when} $k_z = n\uppi$, for $n = \{1, \dots 20\}$.}}
\label{fig:uniform}
\vspace{20pt}
\begin{tikzpicture}
\node[anchor=south west,inner sep=0] at (0,6)
    {\includegraphics[width=\linewidth]{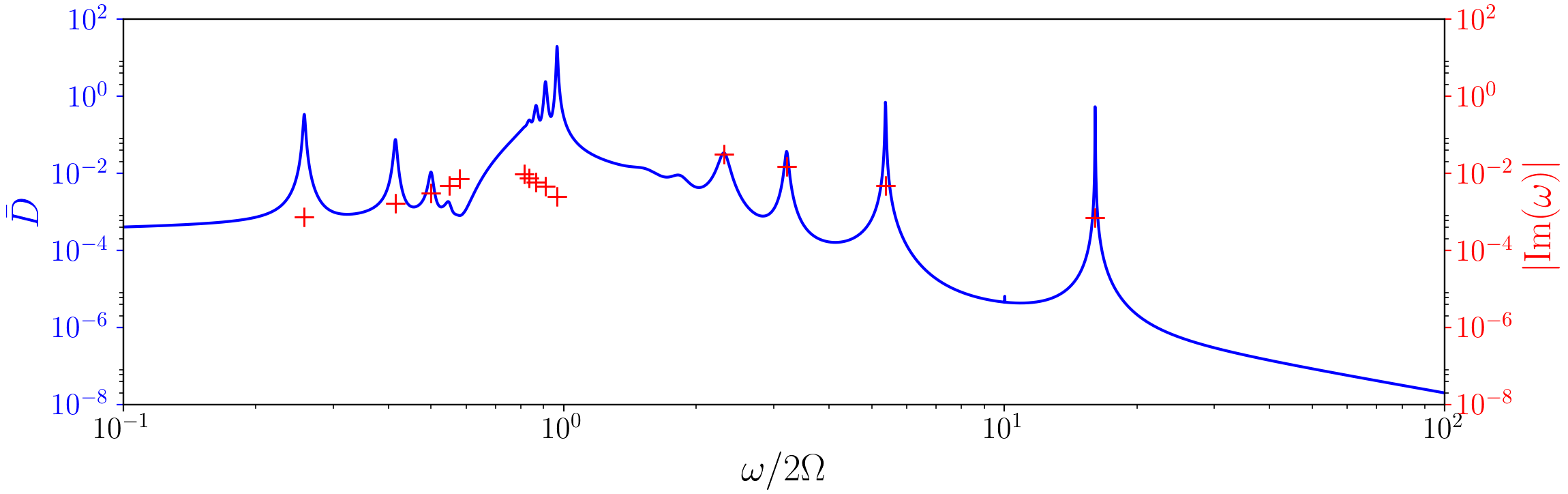}};
\node[anchor=south west,inner sep=0] at (0,0)
    {\includegraphics[width=\linewidth]{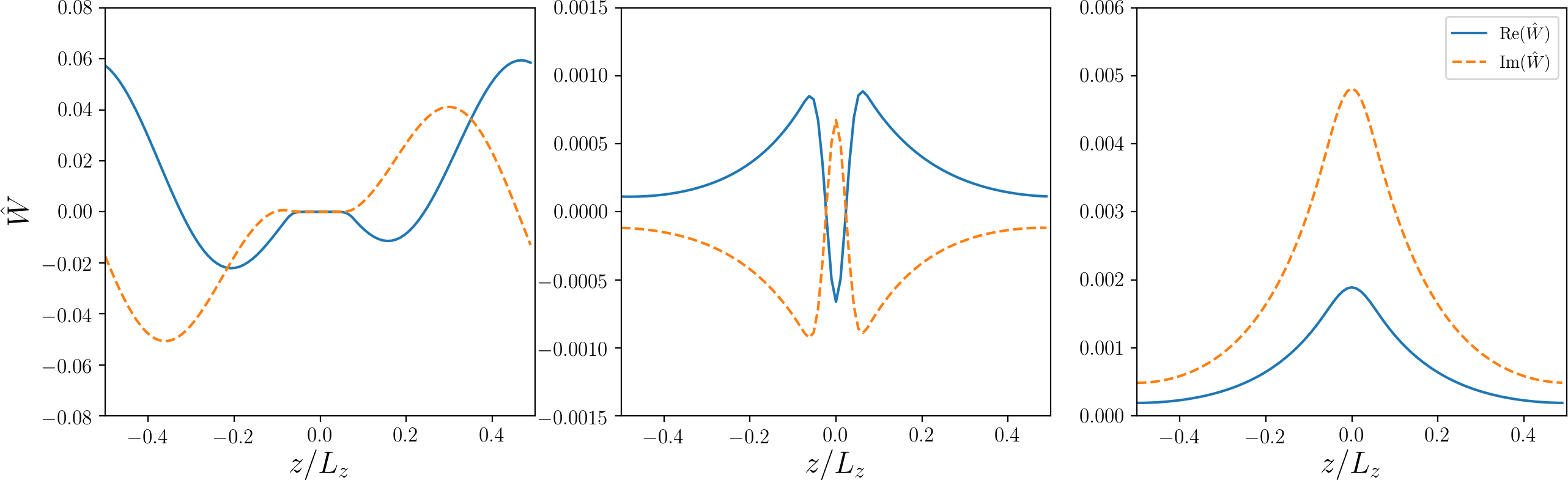}};
\draw[color=black=70, thick, |-|] (1.23,5.8) -- ++ (5.07,0);
\draw[color=black!70, thick, |-|] (7.28,5.8) -- ++ (5.07,0);
\draw[color=black!70, thick, |-|] (13.33,5.8) -- ++ (5.07,0);
\draw[color=black!70, thick, ->] (1.73,5.8) -- ++ (1.7,3.2);
\draw[color=black!70, thick, ->] (7.78,5.8) -- ++ (0.68,3.75);
\draw[color=black!70, thick, ->] (13.83,5.8) -- ++ (-0.885,2.7);
\node[color=black!70] at (2.3,5.25) {\small $\omega/2\Omega \sim 0.258$};
\node[color=black!70] at (2.3+6.05,5.25) {\small $\omega/2\Omega \sim 2.31$};
\node[color=black!70] at (2.3+6.05*2,5.25) {\small $\omega/2\Omega \sim 16.1$};
\end{tikzpicture}\\[-3pt]
\caption{\textit{Top panel:} Dissipation spectrum in our reference case {with periodic boundary conditions and} one step (blue solid line), also displaying the eigenfrequencies (red crosses) of the free Poincar\'e equation, for which the magnitude of the imaginary part is represented as a function of their real part. {\textit{Bottom panel:} $z-$dependence of the vertical velocity, $\hat{W}(z)$, of a short-wavelength inertial eigenmode corresponding to $\omega/2\Omega \sim 0.258$, the leftmost resonance (left panel); of a short-wavelength gravito-inertial eigenmode corresponding to $\omega/2\Omega \sim 2.31$ (middle panel); and of the gravity eigenmode of the staircase, corresponding to $\omega/2\Omega \sim 16.1$, the rightmost resonance (right panel). For each panel, the solid blue line represents the real part of $\hat{W}(z)$ while the dashed orange line represents its imaginary part.}}
\label{fig:eigenfrequencies}
\end{figure*}

On Fig. \ref{fig:uniform}, we show our results for a uniformly stably stratified medium with $N(z)=\bar{N}=10\Omega$, {using periodic (left panel) and rigid (right panel) boundary conditions in the vertical direction. This can be compared with e.g. \cite{AuclairDesrotourEtal2015} in the appropriate cases. In the case of periodic boundary conditions}, we find perfect agreement for the dissipation spectra obtained using the two separate methods, indicating that our code works correctly. The corresponding averaged viscous and thermal dissipation spectra, $\bar{D}_{\text{visc}}$ and $\bar{D}_{\text{ther}}$ respectively, are shown {on both panels of} Fig. \ref{fig:uniform} as a function of the normalised forcing frequency, $\omega/2\Omega$.

\paragraph{{Periodic boundary conditions.}} {On the left panel of Fig. \ref{fig:uniform},} the viscous and thermal dissipation spectra both show a resonant peak centered on the frequency
\begin{equation}
\frac{\tilde{\omega}^{\text{(GIW)}}}{2\Omega} = \left(\bar{S}_{\Omega}^2\frac{k_{\perp}^2}{k^2} + \frac{(\hat{\bm{e}}_{\Omega}\cdot\bm{k})^2}{k^2}\right)^{1/2}
= \sqrt{13.5},
\label{eq:om_giw_}
\end{equation}
which corresponds with the positive root of the dispersion relation for gravito-inertial waves for our chosen parameters. At this particular frequency, the forcing is resonant with {the box-scale} gravito-inertial mode, which is influenced by both rotation and stable density stratification. This leads to enhanced dissipation around that frequency.
We note that given the simple form we have chosen for the forcing, we only obtain one resonant peak. More peaks would be obtained if we were to take a forcing that was a sum over many wavenumbers \citep[][]{OgilvieLin2004,AuclairDesrotourEtal2015}. In a spherical shell, the tidal response is also likely to contain more resonant peaks \citep[e.g.][]{OgilvieLin2004}.

\paragraph{Rigid boundary conditions.} {On the right panel of Fig. \ref{fig:uniform}, we have used rigid boundary conditions in the vertical direction, so that the condition $w(-L_z/2) = w(+L_z/2) = 0$ is imposed. The vertical wave number is then given by
\begin{equation}
k_z^2 = k_{\perp}^2\left(\frac{N^2-\omega^2}{\omega^2-f^2} + \left(\frac{\omega\tilde{f}_{\text{s}}}{\omega^2-f^2}\right)^2\right),
\end{equation}
where $k_{\perp} = 2\uppi$ is imposed by the forcing. We can deduce the corresponding theoretical eigenfrequencies using the quantisation relation
\begin{equation}
k_z = n\uppi,
\end{equation}
where $n$ is an integer \citep{GerkemaShrira2005}. The resulting frequencies have been displayed as grey vertical dashed lines, for $n = \{1, \dots, 20\}.$ For $n = \{1, \dots, 15\}$, these match perfectly the resonant peaks obtained with our numerical code using the Chebyshev method, indicating that this is working correctly. Higher values of $n$ correspond to modes oscillating on smaller scale, which are efficiently damped by diffusive processes. {Note that our forcing does not only excite a single mode in this case, but it instead excites many modes, in general.}
}

\section{The dynamical tide in layered semi-convection}\label{sec:layeredCase}
We now turn to the core of the paper, in which we analyse the forced problem for a layered semi-convective medium. The background buoyancy profile associated with layered semi-convection differs drastically from a uniformly stably stratified or fully convective medium. Accordingly, we expect to obtain different resonances, that are associated with the layered density structure, as we also found in Paper I. This section aims at identifying these, understanding their underlying physics, as well as comparing the dissipation of a density staircase with a fully convective medium, since this is the standard model for giant planet deep interiors.

\subsection{Resonance with free modes of the staircase}\label{freemodes}
Armed with the numerical set up described in Sect. \ref{sec:NumericalSetUp}, we computed spatially-averaged dissipation rates as a function of the forcing frequency. We focus first on the case {with periodic boundary conditions}, with one stably stratified interface in the middle of the box, an aspect ratio $\varepsilon=0.2$, and diffusivity coefficients $\text{E} = K = 10^{-5}$. The dissipation spectrum obtained is displayed on the top panel of Fig. \ref{fig:eigenfrequencies} (blue solid line). We have chosen to only represent the dissipation spectra for positive frequencies, adopting a logarithmic scale to gain clarity, keeping in mind that the dissipation is symmetric with respect to $\omega=0$ \citep[see also][]{OgilvieLin2004,AuclairDesrotourEtal2015} {in this model}.

The {first} feature to note is that the dissipation spectrum contains a number of peaks with enhanced dissipation, which differs from the case of a uniform medium (e.g. left panel of Fig. \ref{fig:uniform}). This illustrates that a region of layered semi-convection possesses a richer set of resonances than a fully convective medium.

To determine the free modes, we set $\bm{F} = \bm{0}$ in Eq. (\ref{eq:forced_Poincare}), to obtain the unforced Poincar\'e equation. This is solved as an eigenvalue problem for the eigenfrequencies and eigenmodes using the same Fourier collocation method. We have done this also for one stably stratified interface with an aspect ratio $\varepsilon=0.2$, and $\text{E}=K=10^{-5}$. The eigenfrequencies are plotted on the top panel of Fig. \ref{fig:eigenfrequencies} as red crosses. The magnitude of the imaginary part of each eigenfrequency (effectively the damping rate of the associated eigenmode) is plotted as a function of its real part (effectively its temporal frequency). We plot only the least damped modes, as these are likely to be the best resolved using our numerical method, and we have discarded certain "junk" eigenmodes that oscillate on the grid-scale.

We see that each peak on the dissipation spectrum corresponds to the frequency of a free mode of the staircase, indicating that their excitation by our forcing is responsible for the peaks. In addition, the narrowest dissipation peaks correspond to the least damped modes, as expected. By lowering the diffusivities, we would expect even more free modes to be excited by the forcing.

We now turn to analyse the spatial structure of the free modes. On the bottom left panel of Fig. \ref{fig:eigenfrequencies}, we have plotted the $z-$dependence of the vertical velocity, $\hat{W}(z)$, of a short-wavelength inertial eigenmode corresponding to the leftmost resonance. This is localised within the convective layer, {(and has $k^{(c)}_z d\approx 2\pi$, using the notation of the later Section 3.2.1)}. On the bottom right panel of Fig. \ref{fig:eigenfrequencies}, we have plotted the $z-$dependence of the vertical velocity, $\hat{W}(z)$, of the gravity eigenmode of the staircase, corresponding to the rightmost resonance. This is primarily localised inside the stably stratified interface. {Finally, on the bottom middle panel of Fig. \ref{fig:eigenfrequencies}, we have plotted the $z-$dependence of the vertical velocity, $\hat{W}(z)$, of a short-wavelength gravito-inertial eigenmode of the staircase, corresponding to $\omega/2\Omega \sim 2.31$.} In each panel, the solid blue line represents the real part of $\hat{W}(z)$ while the dashed orange line represents its imaginary part.

\subsection{Understanding the resonant modes}\label{sec:modes}
In this section, we are interested in understanding {the forced spectral response} of the layered structure. Thus, we choose a forcing amplitude equal to unity, thus independent of frequency.

\begin{figure*}
\centering
\begin{tikzpicture}[scale=0.85]
\node[anchor=south west,inner sep=0] at (0,0)
    {\includegraphics[width=0.85\textwidth]{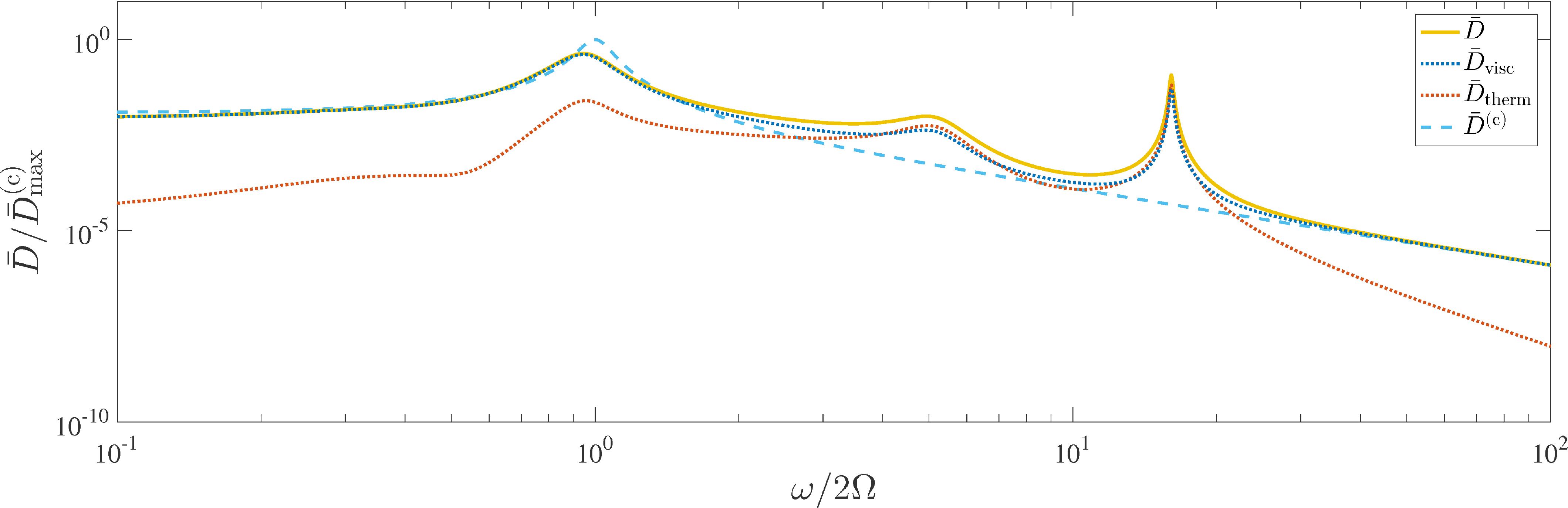}};
\node at (2.5,5.5) {\large $\text{E}=10^{-3}$};
\fill [color=red,opacity=0.08]
       (1.4,1.025) --++ (5.6,0) --++ (0,4.92) --++ (-5.6,0) -- cycle;
\fill [color=purple,opacity=0.08]
       (7,1.025) --++ (5.6,0) --++ (0,4.92) --++ (-5.6,0) -- cycle;
\fill [color=blue,opacity=0.08]
       (12.6,1.025) --++ (5.6,0) --++ (0,4.92) --++ (-5.6,0) -- cycle; 
\end{tikzpicture}\\[-21.9pt]
\begin{tikzpicture}[scale=0.85]
\node[anchor=south west,inner sep=0] at (0,0)
    {\includegraphics[width=0.85\textwidth]{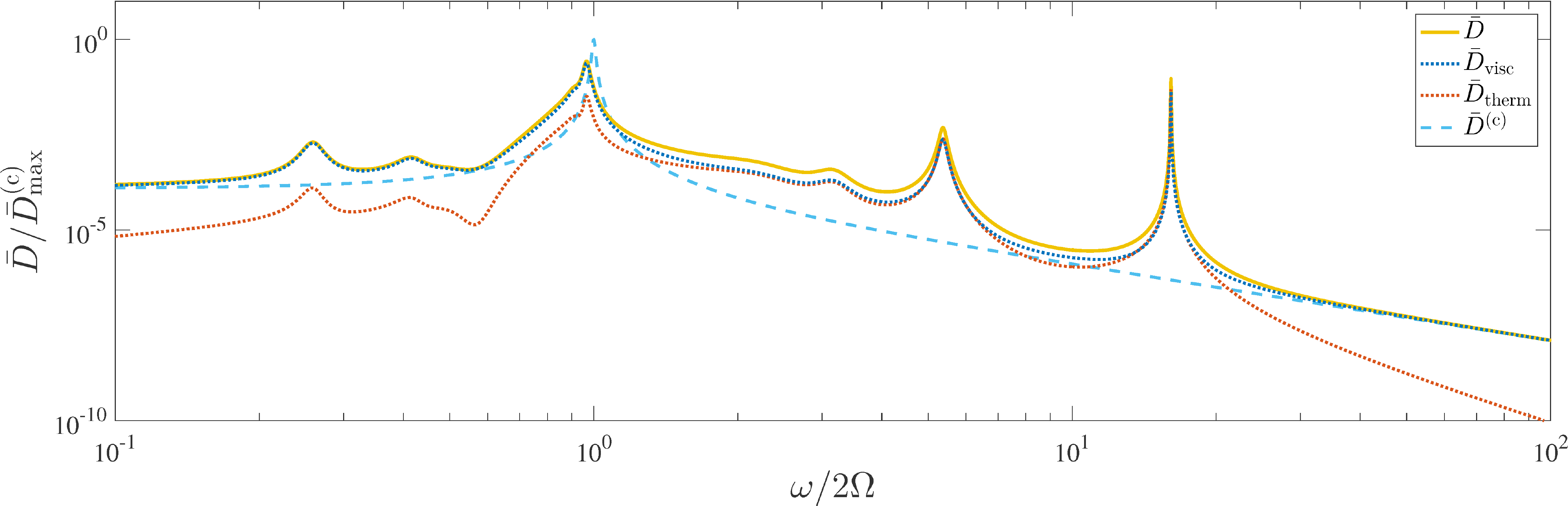}};
\node at (2.5,5.5) {\large $\text{E}=10^{-4}$};
\end{tikzpicture}\\[-21.78pt]
\begin{tikzpicture}[scale=0.85]
\node[anchor=south west,inner sep=0] at (0,0)
    {\includegraphics[width=0.85\textwidth]{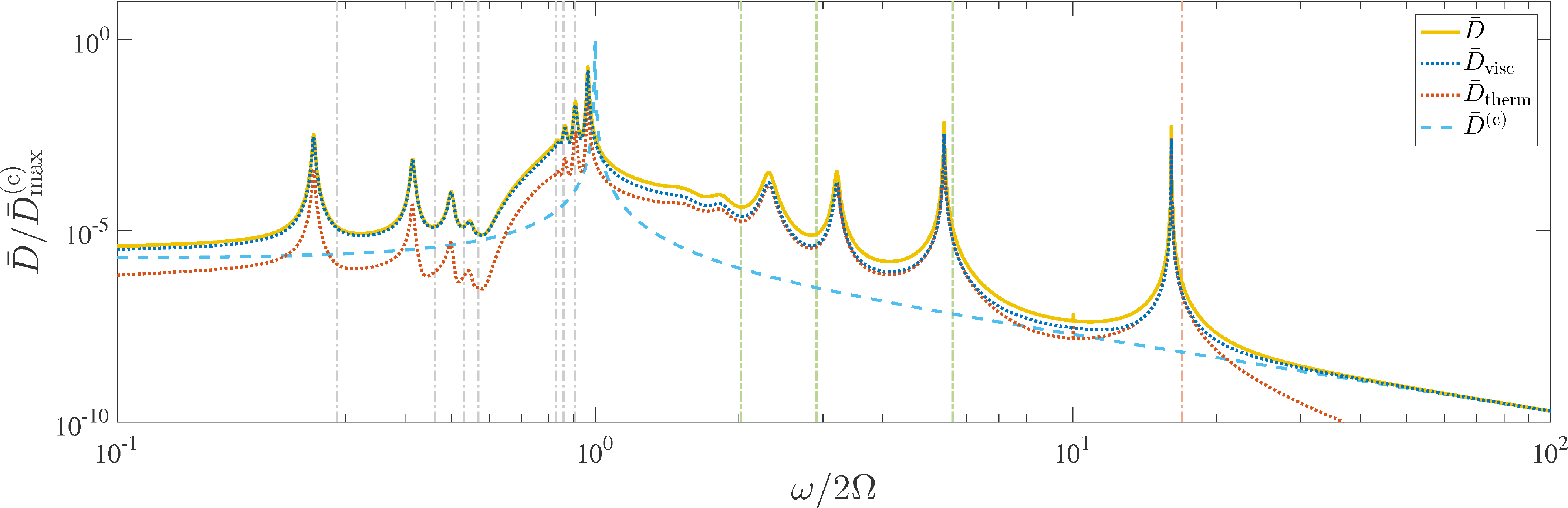}};
\node at (2.5,5.5) {\large $\text{E}=10^{-5}$};
\end{tikzpicture}\\[-5pt]
\caption{Dissipation spectra {with periodic boundary conditions} for a single interface with three different diffusivities (Ekman numbers), $\text{E}=K=10^{-3}$ (top panel), $10^{-4}$ (middle panel) and $10^{-5}$ (bottom panel) and an aspect ratio $\varepsilon=0.2$. The total dissipation is represented by the orange solid line, and its viscous and thermal contributions are represented by the dotted blue and red lines, respectively. The dashed light blue line represents the spatially-averaged dissipation for a fully convective medium, $\bar{D}^{(\text{c})}$. For each panel, the quantity represented is $\bar{D}/\bar{D}^{(\text{c})}_{\text{max}}$, where $\bar{D}^{(\text{c})}_{\text{max}} \equiv \max_{\omega}\bar{D}^{(\text{c})}$, as a function of the normalised frequency $\omega/2\Omega$. In the bottom panel, the vertical dashed-dotted lines indicate the position of characteristic frequencies that are discussed in Section \ref{sec:modes}. Namely, from left to right: resonance with short-wavelength inertial modes (in grey), resonances with short-wavelength super-inertial gravito-inertial modes (in green), and resonance with a free gravity mode of the staircase (in light red).}
\label{fig:spectra}
\vspace{5pt}
\centering
\includegraphics[width=0.9\textwidth]{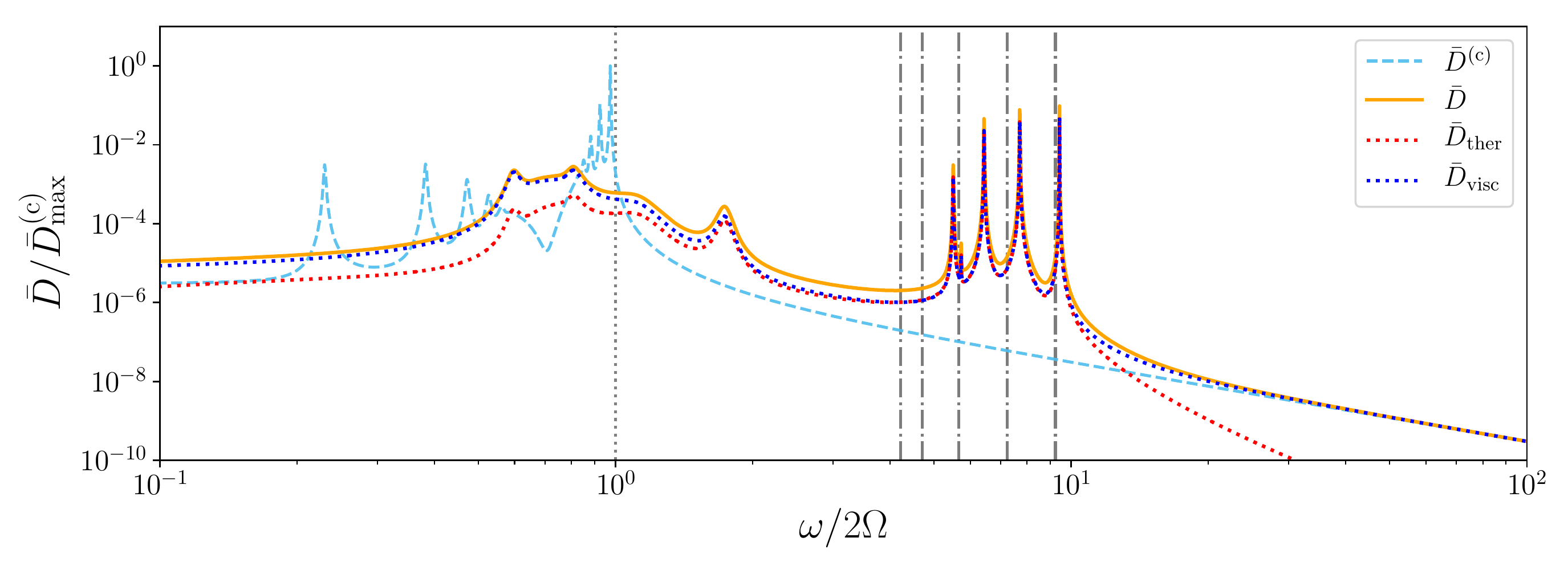}\\[-10pt]
\caption{{Same as the bottom panel of Fig. \ref{fig:spectra} but using rigid boundary conditions in the case with $m=5$ and $\varepsilon=0.5$. Vertical grey dashed dotted lines indicate the five eigenfrequencies calculated from Eq. (\ref{eq:BQFdisprel_finite_NoRotation}). The vertical dotted lined at $\omega=2\Omega$ indicates the upper limit of the sub-inertial frequency range.}}
\label{fig:spectra_rigid}
\end{figure*}

Figure \ref{fig:spectra} shows three dissipation spectra {obtained using periodic boundary conditions}, each computed for one stably stratified interface with $\varepsilon=0.2$. The different panels correspond to decreasing the diffusivities from top to bottom, such that $\text{E} = K = 10^{-3}$, $10^{-4}$ and $10^{-5}$ in the top, middle and bottom panels, respectively. For each panel, the total dissipation rate, $\bar{D}$, is represented by the solid orange line, while its viscous and thermal contributions, $\bar{D}_{\text{visc}}$ and $\bar{D}_{\text{ther}}$, are represented by the dotted blue and red lines, respectively. For comparison, the dissipation spectrum in a fully convective medium, $\bar{D}^{(\text{c})}$, is represented by the dashed light blue line, and all the dissipation rates have been normalised by $\bar{D}^{(\text{c})}_{\text{max}} \equiv \max_{\omega}\bar{D}^{(\text{c})}$.

In agreement with \cite{OgilvieLin2004} and \cite{AuclairDesrotourEtal2015}, the resonant peaks are more numerous and narrower when the viscosity (and thermal diffusivity) is decreased to reach the smaller values that are more relevant to planetary or stellar interiors. However, while our choice of parameters would give only one resonant peak in a uniformly stably stratified or fully convective medium -- centered on $\omega/2\Omega = 1$ in the latter case (see Fig. \ref{fig:spectra}) -- we clearly see that the layered structure introduces new resonances. {As a result, it is clear that for $\text{E}=K \leq 10^{-4}$ (see the two bottom panels of Fig. \ref{fig:spectra}), the total dissipation in the layered case is higher than in the convective case, except in a narrow frequency window around the Coriolis frequency, $2\Omega$. This discrepency also seems to become more important when diffusivities get smaller, especially near the resonances introduced by the layered structure.} Another observation is that these additional resonances are broadly distributed over the frequency spectrum. Some correspond to resonances with inertial modes, corresponding to frequencies $\omega \lesssim 2\Omega$ (red region on top panel of Fig. \ref{fig:spectra}); some with {super-inertial} gravito-inertial modes, corresponding to frequencies $2\Omega \lesssim \omega \lesssim \bar{N}$ (purple region on top panel of Fig. \ref{fig:spectra}); and finally some with gravity modes, corresponding to frequencies $\bar{N} \lesssim \omega \lesssim N_0$ (blue region on top panel of Fig. \ref{fig:spectra}).

In what follows, we identify the underlying physics behind these resonances, building upon Section \ref{freemodes}. On the bottom panel of Fig. \ref{fig:spectra}, corresponding to the lowest value of the diffusivities ($\text{E}= K = 10^{-5}$), we have indicated by vertical dashed-dotted lines the position of particular frequencies. These frequencies are described by simple dispersion relations, that were also found in \cite{ABM2017} to correspond with waves that are efficiently transmitted across a density staircase. These are found to be good candidates to explain the resonances that are observed, and are described in further detail below.

\subsubsection{Resonance with short wavelength inertial modes}
In the inertial regime ($\omega<2\Omega$), a succession of resonances with enhanced dissipation appear as we decrease the viscosity. These correspond to inertial modes with vertical semi-wavelengths that fit inside the convective layer. The gray dashed lines on the bottom panel of Fig. \ref{fig:spectra} thus correspond to frequencies that obey the relation $\lambda_z/2 = nd$, or equivalently 
\begin{equation}
k_z^{(\text{c})}(\omega)d = n\uppi,
\label{eq:kzdceqnpi}
\end{equation}
for different integers $n$. Here, $d$ is the vertical extent of the convective region (see Fig. \ref{fig:Nprofile}), and to draw the vertical lines on Fig. \ref{fig:spectra}, we use the vertical wavenumber in the adiabatic limit,
\begin{equation}
k_z^{(\text{c})} = k_{\perp}\left[\frac{\omega^2(f^2+\tilde{f}_{\text{s}}^2-\omega^2)}{(\omega^2-f^2)^2}\right]^{1/2},
\label{eq:kz}
\end{equation}
where the label `$^{\text{(c)}}$' is used to stress that this is the vertical wave number in a convective medium. {We recall that the expressions of the quantities $f$ and $\tilde{f}_{\text{s}}$ are given in Section \ref{subsec:MainAssumptions}}. The expression above is obtained from the dispersion relation of pure inertial waves,
\begin{equation}
\omega^2 = \frac{(2\bm{\Omega}\cdot\bm{k})^2}{k_{\perp}^2+k_z^{\text{(c)2}}},
\label{eq:disprelIW}
\end{equation}
where $\bm{k} = k_{\perp}\hat{\bm{e}}_{\chi} + k_z^{\text{(c)}} \hat{\bm{e}}_z$. 

The discrepancy between the predicted and actual positions of those resonances can be partly explained twofold. Firstly, the vertical wavenumber above corresponds to the adiabatic case. Secondly, the stably stratified interface in the middle of the box has a non-negligible vertical extent in which even pure inertial waves become influenced by buoyancy. This differs from the idealised model that was used in \citetalias{ABM2017}, in which stably stratified interfaces were infinitesimally thin. 
We recall that in \citetalias{ABM2017}, the modes matching the condition given by Eq. (\ref{eq:kzdceqnpi}) were also found to be efficiently transmitted through a density staircase.

\subsubsection{Resonance with short wavelength super-inertial gravito-inertial modes}
Based on a similar idea, we have also looked for modes in the gravito-inertial regime with vertical semi-wavelengths that fit inside the thin stably stratified interfaces, in which they are propagative. On the bottom panel of Fig. \ref{fig:spectra}, the dotted-dashed green vertical lines correspond to frequencies such that
\begin{equation}
k_z(\omega)l = n\uppi
\label{eq:kzdeqnpi}
\end{equation}
for three different integers $n = \{1,2,3\}$. The frequency is given by the dispersion relation of gravito-inertial waves,
\begin{equation}
\omega^2 = N_0^2\frac{k_{\perp}^2}{k_{\perp}^2+k_z^2} + \frac{(2\bm{\Omega}\cdot \bm{k})^2}{k_{\perp}^2+k_z^2},
\label{eq:GIWs}
\end{equation}
where $k_z$ is now the vertical wavenumber in a stably stratified medium, characterised by the constant buoyancy frequency $N_0$. We note that these waves are evanescent in the convective layer since $2\Omega < \omega < N_0$ \cite[see][]{MathisNeinerTranminh2014}.

Since the buoyancy frequency is not constant in the stably stratified region (see Fig. \ref{fig:Nprofile}), $N_0$ was chosen to match the corresponding peaks as closely as possible. Its value was restricted so that $\bar{N} < N_0 < \max_z N(z)$. However, we note that we do not expect the vertical lines to perfectly match the position of the resonant peaks because $N^2(z)$ is not constant in the interfaces, {and additionally  the dispersion relation used above (Eq. \ref{eq:GIWs}) was obtained in the adiabatic case.} Nevertheless, this clearly explains the physics of these resonances with enhanced dissipation.

\subsubsection{Resonance with the free gravity mode of the staircase}\label{sec:BQF}
In \citetalias{ABM2017}, we derived the following dispersion relation for the free modes of the staircase (extending prior work by \cite{BelyaevQuataertFuller2015}):
\begin{equation}
\omega^2 = \bar{N}^2 \left(
\frac{(\bar{k}d)^2/k_zd}{2\coth (k_zd)-2\cos\theta\csch (k_zd)}
\right),
\label{eq:BQFdisprel_finite}
\end{equation}
where the stably stratified interfaces were modelled as discontinuous jumps.
Here, $k_z$ is given by Eq. (\ref{eq:GIWs}), $\bar{k}=k_{\perp}\omega/\sqrt{(\omega^2-f^2)}$, and $\cos\theta$ is one of the roots of the polynomial
\begin{equation}
T_m(\cos\theta) + [\cos\theta \coth (k_zd) - \csch (k_zd)]U_{m-1}(\cos\theta) = 0,
\end{equation}
where $T_m$ and $U_m$ are Chebyshev polynomials of the first and second kinds, respectively, with their order $m$ being equal to the number of convective steps.

Since $k_z$ has a complex dependence on $\omega$, it is challenging to extract the roots of the dispersion relation given by Eq. (\ref{eq:BQFdisprel_finite}) in general. However, when looking at gravity modes, the effect of rotation can be neglected, so that Eq. (\ref{eq:BQFdisprel_finite}) reduces to
\begin{equation}
\omega^2 = \bar{N}^2 \left(
\frac{k_{\perp}d}{2\coth (k_{\perp}d)-2\cos\theta\csch (k_{\perp}d)}
\right)
\label{eq:BQFdisprel_finite_NoRotation}
\end{equation}
\citep{BelyaevQuataertFuller2015}. This particular frequency (when $m=1$) is displayed as the dotted-dashed light red vertical line on the bottom panel of Fig. \ref{fig:spectra}. Its position matches rather well the position of the rightmost resonant peak, though we do not expect a perfect match because we have neglected the effects of rotation and diffusion, and we have modelled the interfaces as smooth rather than discontinuous jumps, unlike the assumptions that went into the derivation of this expression.

The dispersion relation above has exactly $m$ roots, so that there could be up to $m$ resonant peaks corresponding to resonances with free modes of the staircase, visible on the dissipation spectra. We retrieve this property when using rigid boundary conditions with more than one step. {This is shown on Fig. \ref{fig:spectra_rigid}, which displays the same quantities as on the bottom panel of Fig. \ref{fig:spectra} but using rigid boundary conditions in the case with $m=5$ and $\varepsilon=0.5$. The five roots of the dispersion relation written above (Eq. (\ref{eq:BQFdisprel_finite_NoRotation})) are overplotted as grey vertical dashed dotted lines. They approximately match the five resonant peaks on the dissipation spectra (these frequencies were found to match more closely than the free modes of a uniformly stratified medium with rigid conditions). The discrepancy can be explained{,as in the previous case,} by noting that Eq. (\ref{eq:BQFdisprel_finite_NoRotation}) has been derived in a non-rotating plane-parallel model, assuming infinitesimally thin stably-stratified interfaces. This differs from our numerical setup which includes rotation, and contains stably stratified layers with a finite size for numerical reason.}

\subsubsection{Changing the aspect ratio and the number of steps}
\begin{figure*}
\centering
\begin{tikzpicture}
\node[anchor=south west,inner sep=0] at (0,0)
	{\includegraphics[width=0.85\linewidth]{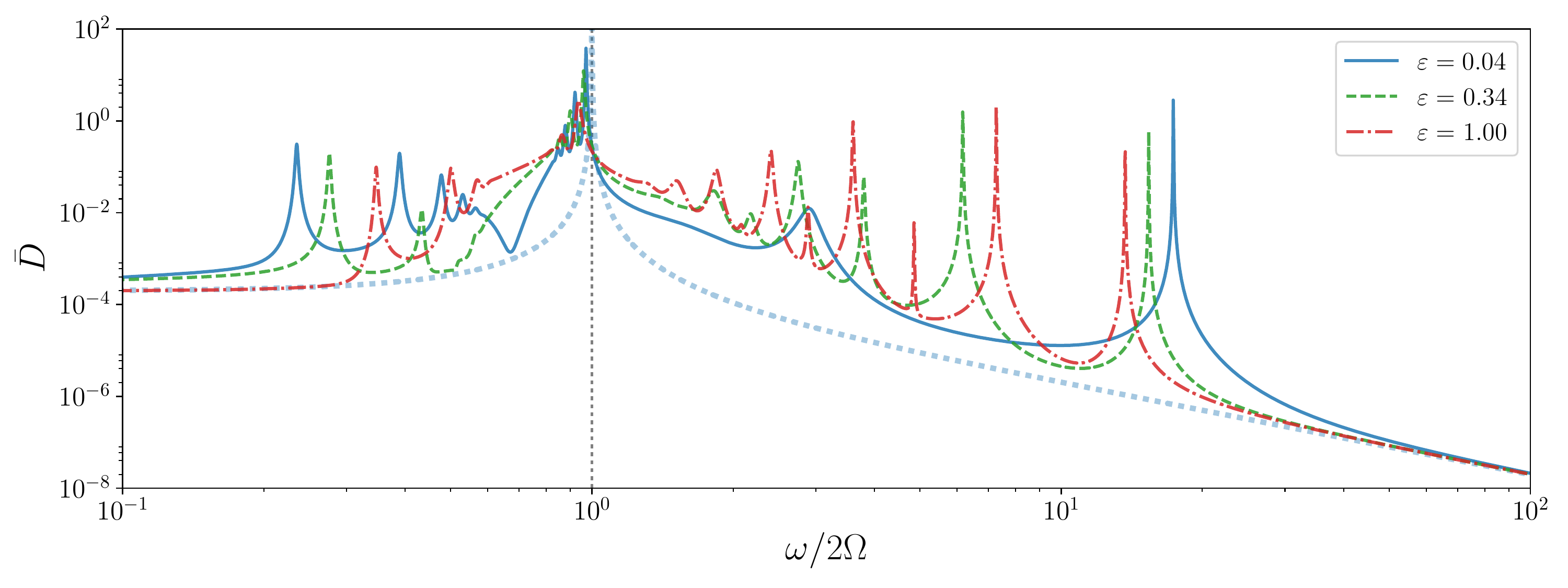}};
    \draw[fill=white,white] (1,0) rectangle (15.45,0.95);
\end{tikzpicture}\\[-26pt]
\includegraphics[width=0.85\linewidth]{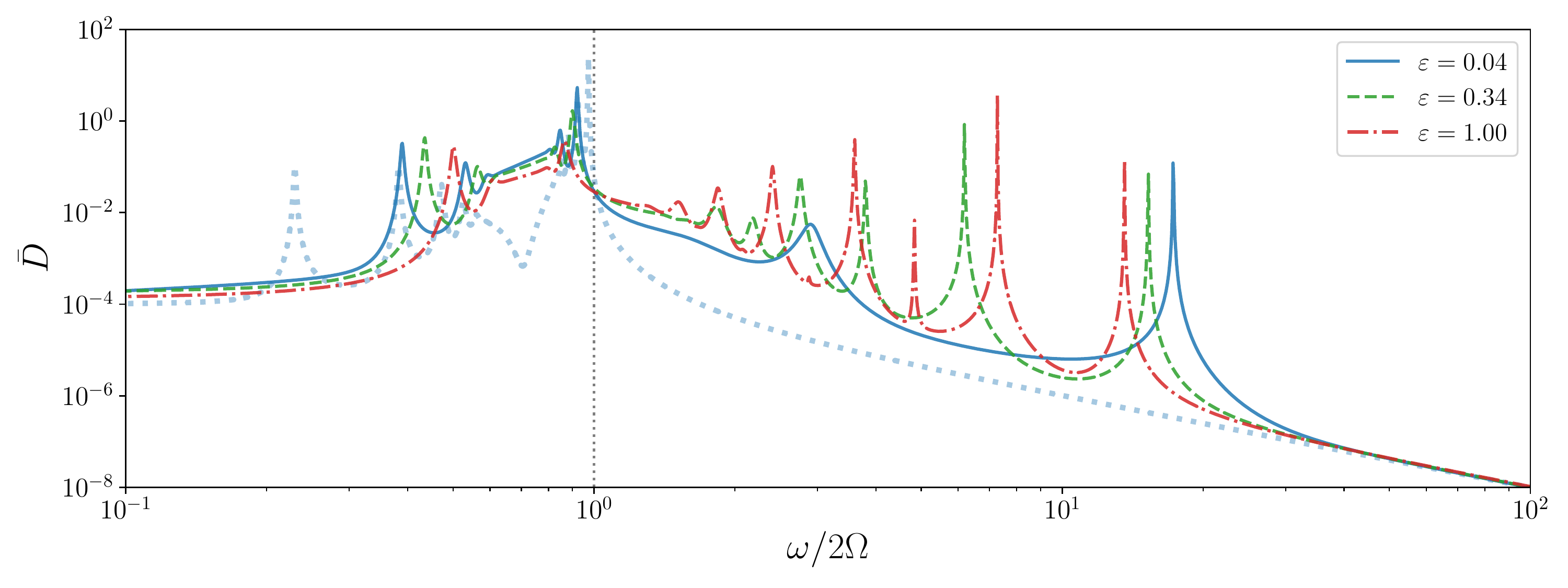}\\[-8pt]
\caption{Dissipation spectra obtained with $m=1$ step for the different aspect ratios $\varepsilon=0.04$ (blue solid line), 0.34 (green dashed line), and 1.00 (red dashed dotted line) and $\text{E}=K=10^{-5}${, using vertically periodic boundary conditions \textit{(top panel)} and vertically rigid boundary conditions \textit{(bottom panel)}}. {For comparison, the dotted light blue line corresponds to the case of a fully convective medium.}}
\label{fig:compspectra_eps}
\centering
\begin{tikzpicture}
\node[anchor=south west,inner sep=0] at (0,0)
	{\includegraphics[width=0.85\linewidth]{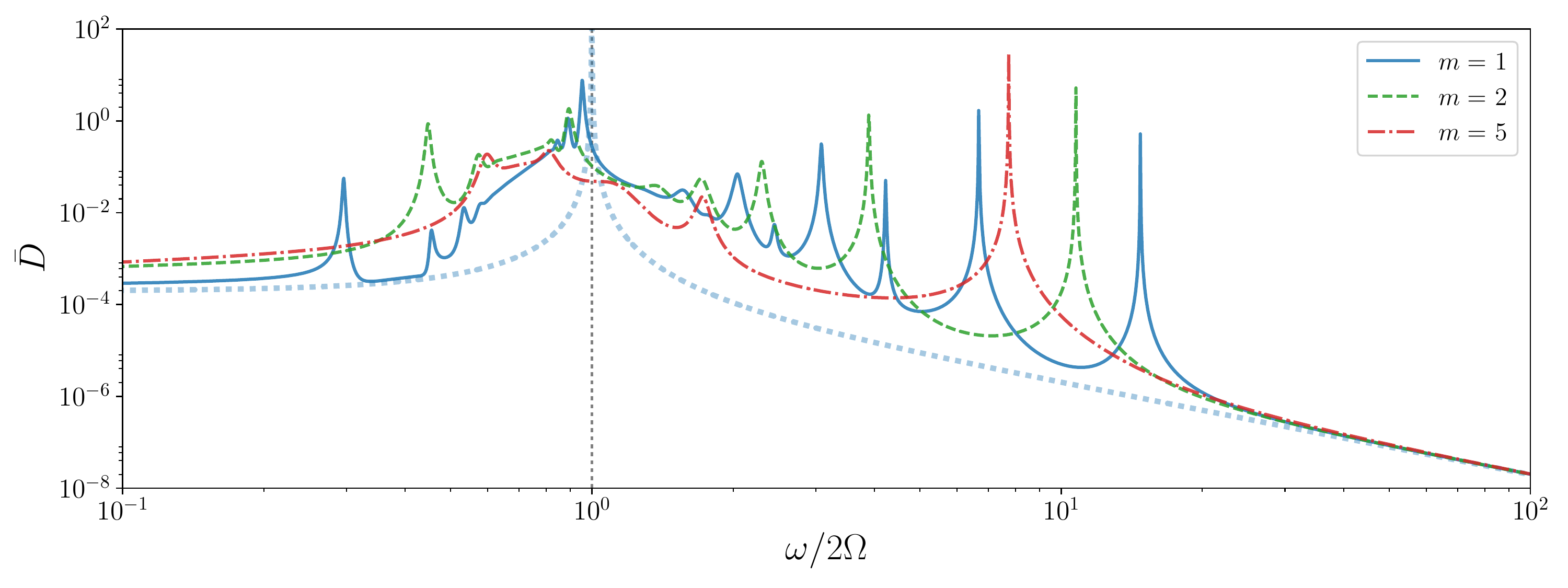}};
\draw[fill=white,white] (1,0) rectangle (15.45,0.95);
\end{tikzpicture}\\[-26pt]
\includegraphics[width=0.85\linewidth]{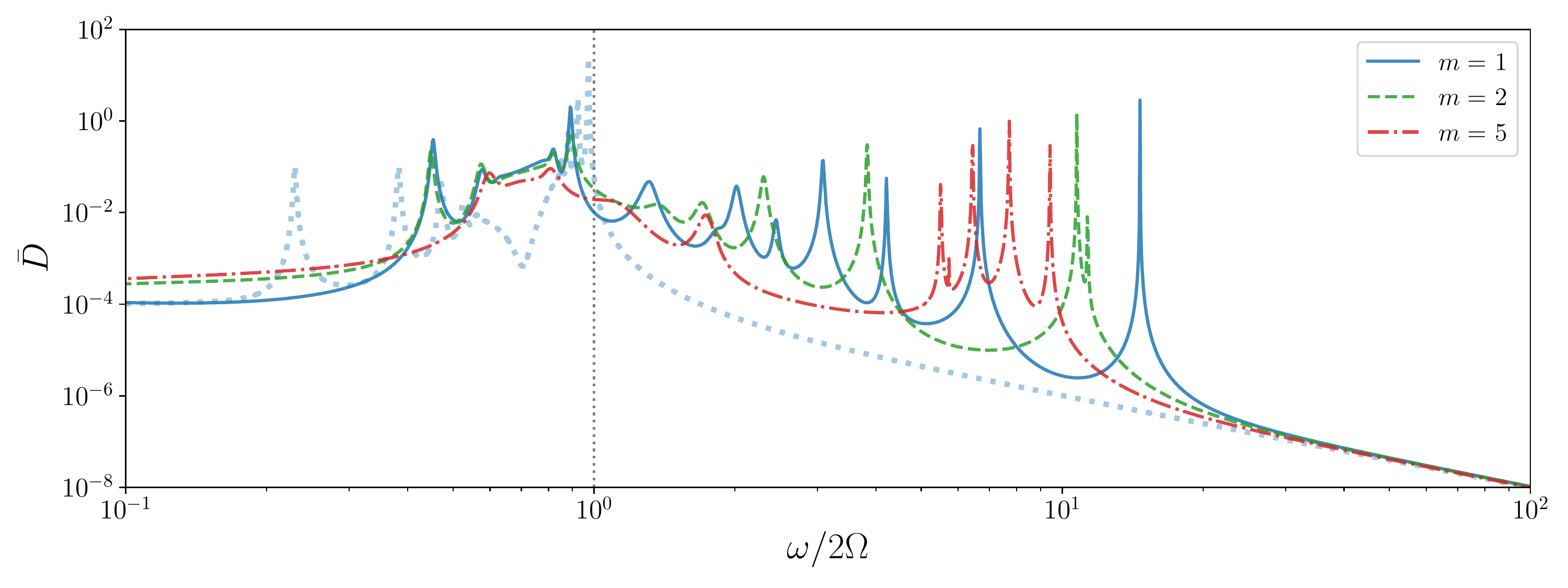}\\[-8pt]
\caption{Dissipation spectra obtained with $\varepsilon=0.5$ for different numbers of steps $m=1$ (blue solid line), 2 (green dashed line), and 5 (red dashed dotted) and $\text{E}=K=10^{-5}${, using vertically periodic boundary conditions \textit{(top panel)} and vertically rigid boundary conditions \textit{(bottom panel)}}. {For comparison, the dotted light blue line corresponds to the case of a fully convective medium.}}
\label{fig:compspectra_m}
\end{figure*}

Given the physical interpretation in the previous sections, we expect the location of the resonant peaks to change with the size of the convective layers (for resonances with short wavelength sub-inertial modes) and the size of the stably stratified layers (for resonances with short wavelength super-inertial gravito-inertial modes). {We also expect this to be true for resonances with free modes of the staircase, since the corresponding dispersion relation, given by Eq. (\ref{eq:BQFdisprel_finite_NoRotation}), explicitly depends on the size of the convective layers $d$ and the number of stably stratified interfaces $m$.}
{To illustrate this, we show how dissipation spectra are modified when we vary the aspect ratio $\varepsilon$, and the number of stably stratified interfaces $m$.}

\paragraph{Changing the aspect ratio.} {Figure} \ref{fig:compspectra_eps} shows three different dissipation spectra corresponding to the aspect ratios $\varepsilon=0.04$ (blue solid line), 0.34 (green dashed line), and 1.00 (red dashed dotted line), all with one step and $\text{E}=K=10^{-5}$. {The top panel corresponds to periodic boundary conditions, while the bottom panel corresponds to rigid ones.}

Let us first focus on the viscous dissipation rate obtained in a fully convective medium (light blue lines). In the case with periodic boundary conditions (top panel), a single resonant peak is observed at $\omega = 2\Omega$, while in the case with rigid boundary conditions (bottom panel), the forcing imperfectly excites a set of short wavelength inertial modes. This is similar to what we have identified in the case of a uniformly-stratified medium in Section \ref{sec:uniform}.

Let us consider the dissipation spectra corresponding to the layered case. In the sub-inertial frequency range, resonant peaks are more numerous for small aspect ratios, while in the super-inertial frequency range, resonant peaks are more numerous for large aspect ratios (since larger wavelength modes can then fit inside the stably stratified layer). In addition, the rightmost resonant peak shifts to smaller frequencies as the aspect ratio increases. This is consistent with Eq. (\ref{eq:BQFdisprel_finite_NoRotation}), which tells us that $\omega$ is an increasing function of $d$ (the size of the convective layer).

Let us now compare the differences between boundary conditions, i.e. between both panels of Fig. \ref{fig:compspectra_eps}. First, it can be seen that the dependence of the main features of the spectra in term of number of resonant peaks, their shapes, and their associated frequencies, is qualitatively similar with both sets of boundary conditions. However, we find some differences in the sub-inertial range: for the same aspect ratio, more resonances with short wavelength inertial modes are observed in the case with periodic boundary conditions.

\paragraph{Changing the number of steps.} Figure \ref{fig:compspectra_m} shows three different dissipation spectra corresponding to different numbers of steps $m=1$ (blue solid line), 2 (green dashed line), and 5 (red dashed dotted line), again all with $\text{E}=K=10^{-5}$. In order to keep the volume fraction occupied by the stably stratified layers constant, we kept the aspect ratio constant ($\varepsilon=0.5$ here) while changing the value of $m$. {We note that} changing the number of steps while keeping the length scale over which the forcing term varies (the size of the box), means changing the relative scale between the size of the steps and the forcing. {The top panel corresponds to periodic boundary conditions while the bottom panel corresponds to rigid ones. On both panels, the viscous dissipation rate in a fully convective medium is displayed as the dotted light blue line.}

The resonances with short wavelength inertial waves (in the convective layers), and with sub-inertial gravito-inertial waves (in the stably stratified interfaces), both become weaker and less numerous as we increase the number of steps (see Fig. \ref{fig:compspectra_m}). This is partly because the forcing that we have adopted varies on the box-scale, so it will most efficiently excite  waves that vary on this length-scale, and it will excite the shorter-wavelength waves much less efficiently. It is also partly due to the increasing damping efficiency of viscosity and thermal diffusion for these short-wavelength waves.

Let us now compare the differences between boundary conditions, i.e. between both panels of Fig. \ref{fig:compspectra_m}. First, we find the same differences in the sub-inertial range as mentioned in the paragraph above. In the gravito-inertial frequency range, the strength and number of the resonant peaks that are observed is qualitatively similar. However, in the frequency range of gravity modes, the following major differences arise. 
In the case with periodic boundary conditions (top panel of Fig. \ref{fig:compspectra_m}), the forcing only excites one global mode of the staircase, that is on the box scale. Thus, only one resonant peak corresponds to a resonance with a free gravity mode of the staircase. {We note that the corresponding} frequency is shifted to lower frequencies when the number of steps is increased. This is because the size of the convective steps, $d$, then gets smaller (see discussion above). On the other hand, in the case of rigid boundary conditions, (bottom panel of Fig. \ref{fig:compspectra_m}), the forcing excites exactly $m$ resonances, that correspond to the $m$ free modes of the staircase described in Section \ref{sec:BQF}. Understanding why these $m$ resonances do not appear separately when adopting vertically periodic boundary conditions requires further investigation, but we think this is unlikely to be the case in a more realistic calculation.

\subsection{Exploration of frequency-averaged dissipation rates in parameter space}\label{sec:ParameterSpace}
\begin{figure*}
\centering
\includegraphics[height=5.4cm]{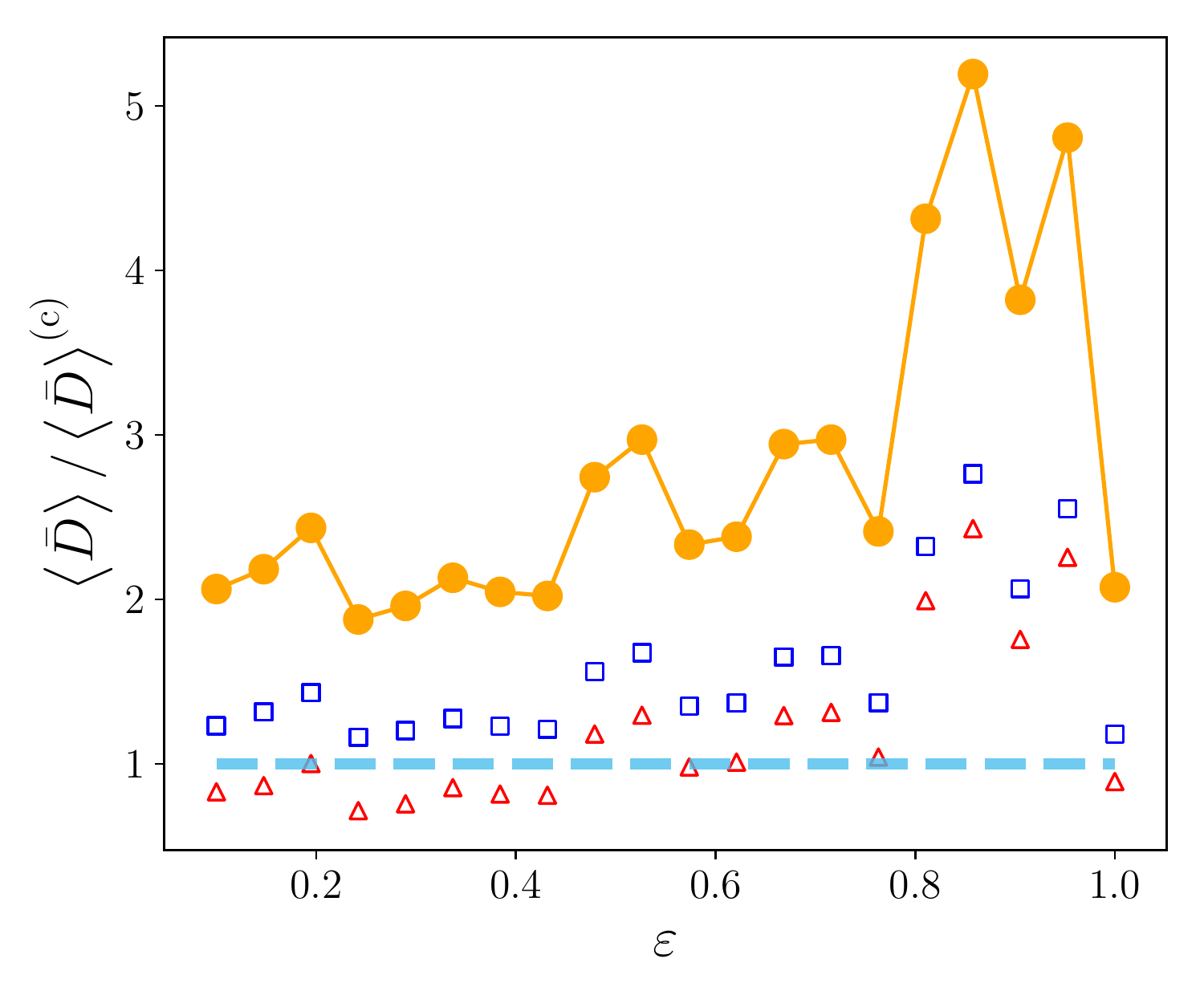}
\hspace{-5pt}
\includegraphics[height=5.4cm]{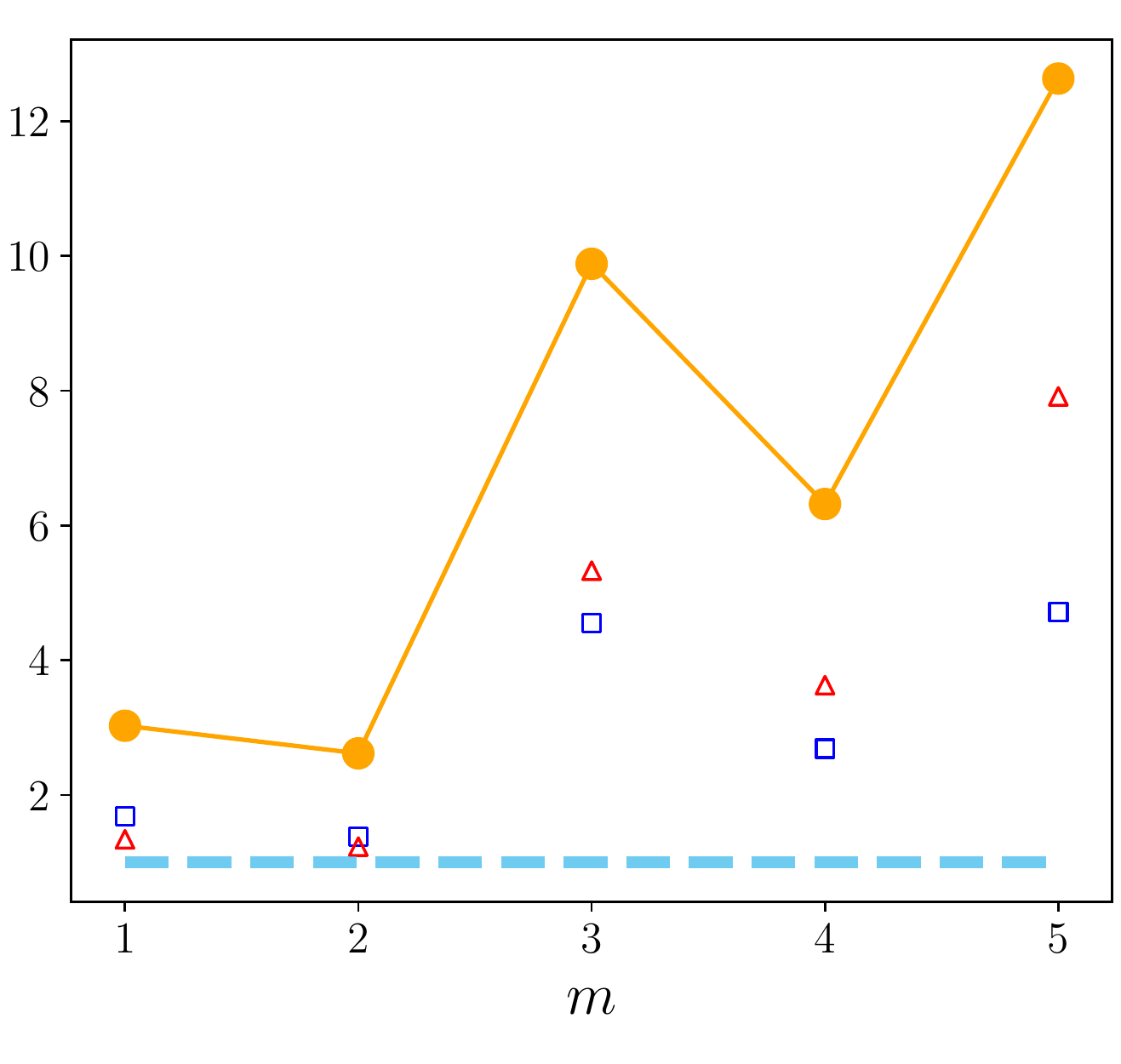}\hspace{2pt}
\includegraphics[height=5.4cm]{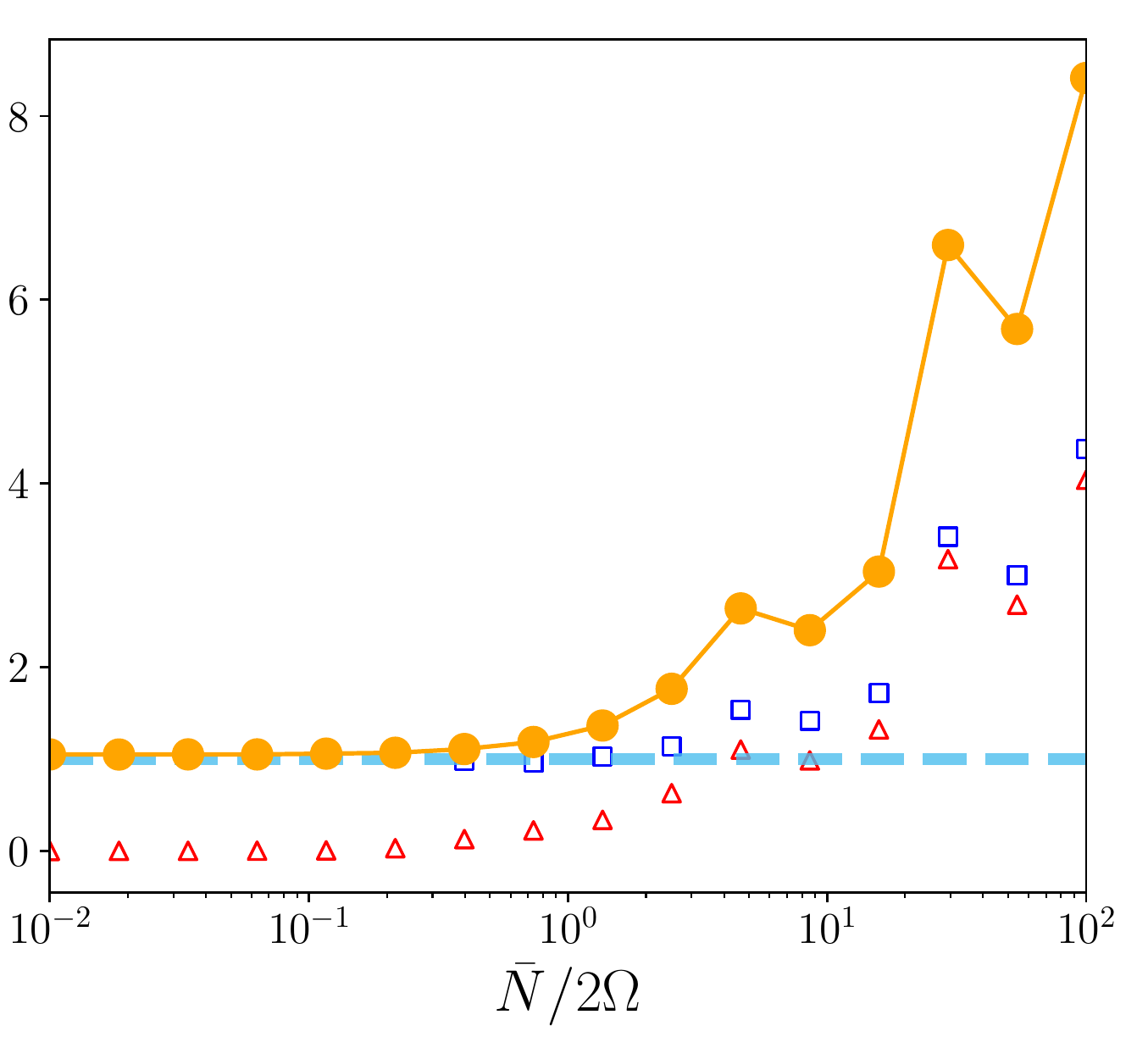}\\[-10pt]
\caption{Frequency-averaged dissipation rates as a function of the aspect ratio (\textit{left panel}), of the number of steps (\textit{middle panel}), and of the ratio $\bar{S}_{\Omega} = \bar{N}/2\Omega$ (\textit{right panel}), using vertically periodic boundary conditions. These results were obtained for one stably stratified interface and $\text{E}=K=10^{-5}$. For each panel, the orange filled circles correspond to the total dissipation rate, while its viscous and thermal contributions are represented by the empty red triangles and blue squares, respectively. On the $y-$axis, the frequency-averaged dissipation rates in the layered case, $\bar{\left<D\right>}$, is normalised by the frequency-averaged dissipation rate of the fully convective case, $\bar{\left<D\right>}^{\text{(c)}}$, which is also indicated by the horizontal blue dashed line. We adopted a forcing amplitude that is proportional to $\omega$ for these calculations.}
\label{fig:nu-5_Eps}
\vspace{8pt}
\centering
\includegraphics[height=5.4cm]{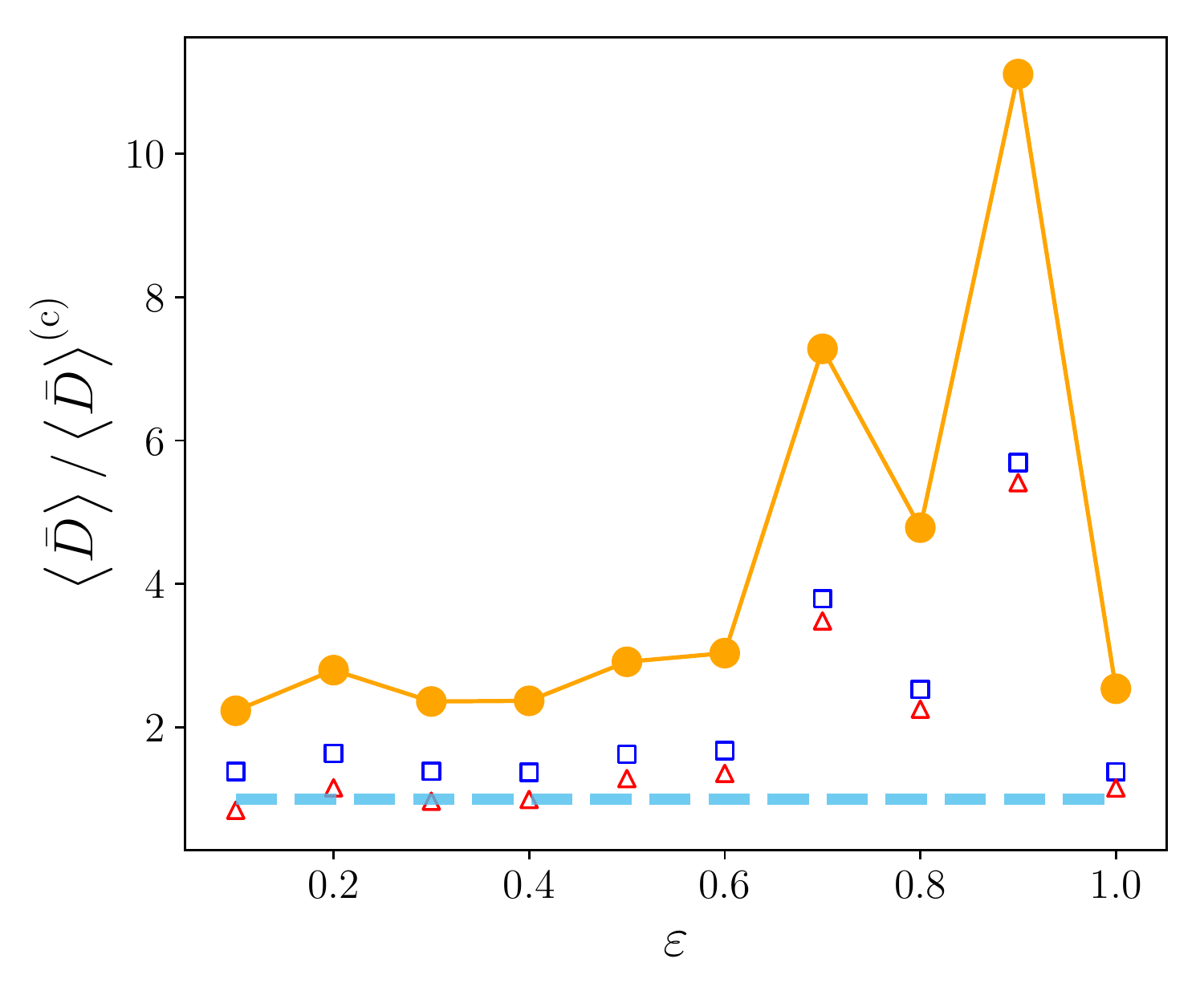}
\hspace{-2pt}
\includegraphics[height=5.4cm]{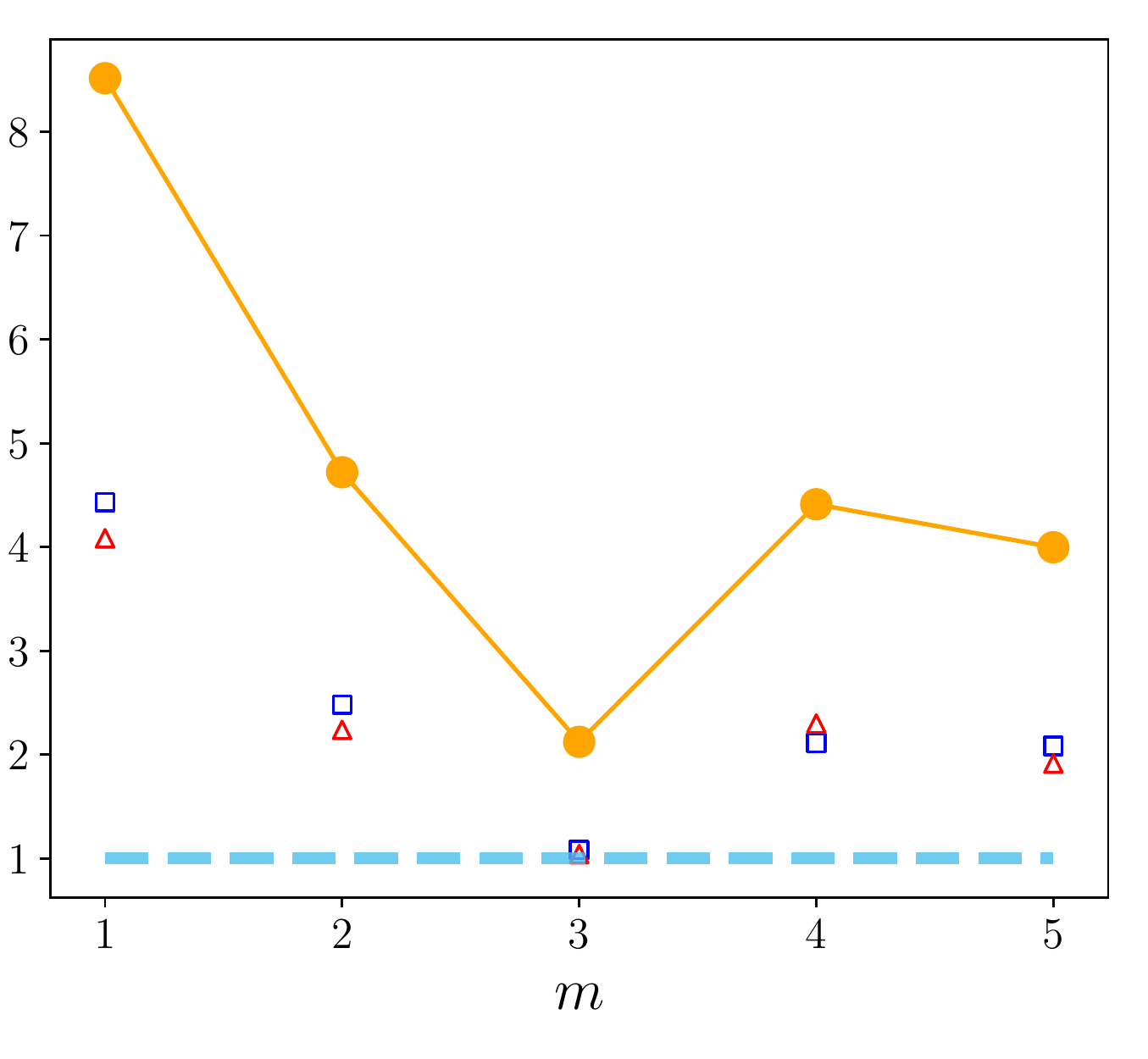}\hspace{-4pt}
\includegraphics[height=5.4cm]{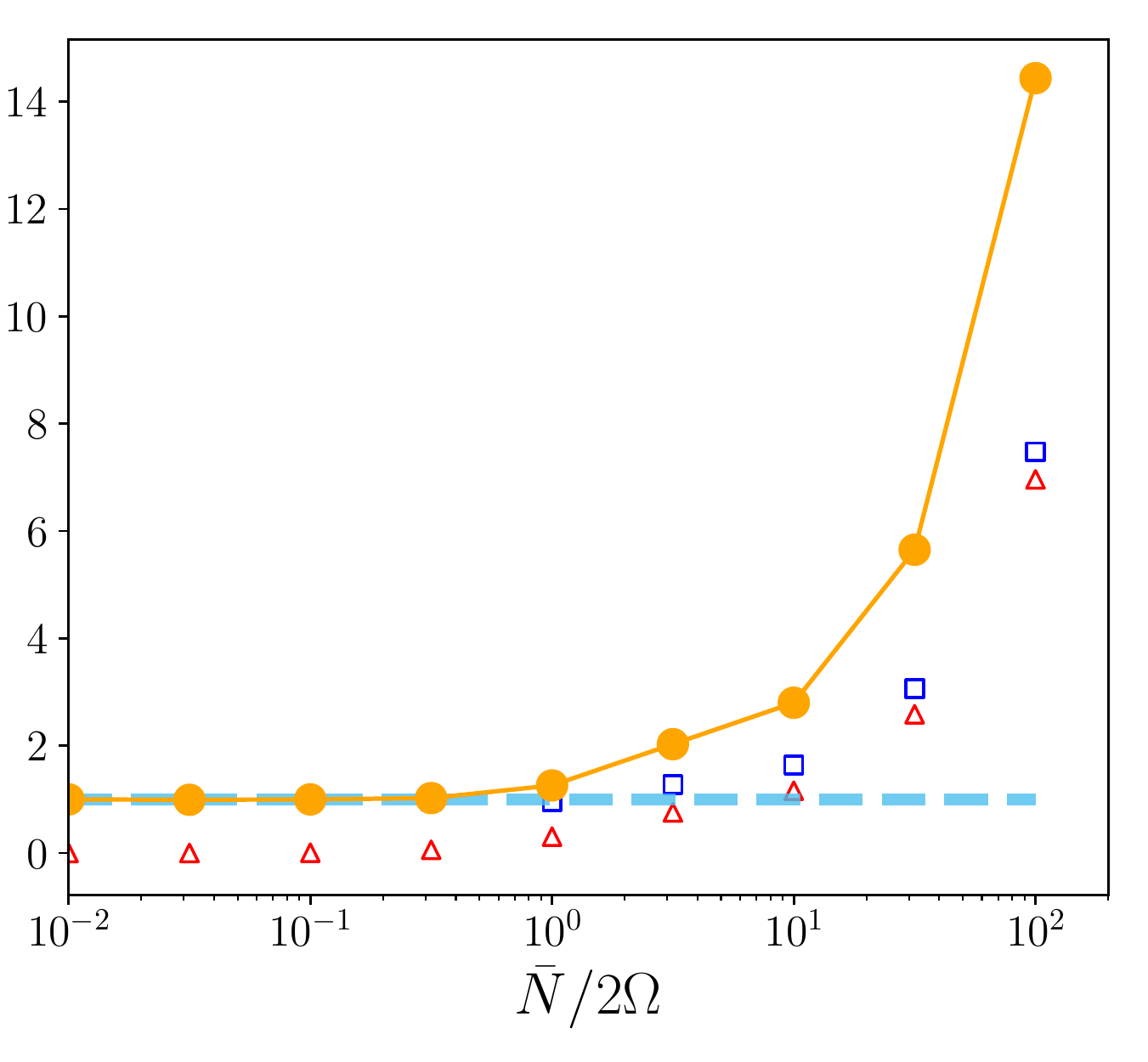}\\[-10pt]
\caption{Same as Fig. \ref{fig:nu-5_Eps}, but using vertically rigid boundary conditions.}
\label{fig:nu-5_Eps_ref}
\end{figure*}
We have so far explored the response of a layered density structure to an imposed forcing, and we have given a physical interpretation for each of the corresponding resonances with enhanced dissipation.
Now, we would like to explore how the resulting dissipation in a region of layered semi-convection compares with that in a fully convective medium. In this section, we thus compute frequency-averaged dissipation rates with a forcing amplitude that is proportional to the tidal frequency $\omega$. 
This is a more realistic model for the excitation of tidal waves, because the amplitude of the forcing term, which drives tidal waves, contains two terms that scale as $2\Omega\, \omega$ and $\omega^2$, respectively \citep[see e.g.][and discussion in Section \ref{sec:forcing}]{Ogilvie2005,Ogilvie2014}. We choose here to keep the one that is dominant in the regime of low tidal frequencies. We keep in mind that the term in $\omega^2$ enhances the contribution of resonances that are excited at higher frequencies, namely resonances with short wavelength gravito-inertial modes, and gravity-like modes (that are excited at frequencies $\omega > 2\Omega$, see Section \ref{sec:modes}). These are specific to the layered case. We note that we have explored calculations for which the forcing scales as $\omega^2$ using a cutoff frequency to cut the non-resonant part of the spectra. These gave the same qualitative trends that for an $\omega$ forcing described hereafter.

{In order to unravel the influence of the layered structure alone, we successively focus on three quantities that parametrise the layered structure: $(\varepsilon,m,\bar{N}/2\Omega)$.}
We {then} compute frequency-averaged dissipation rates according to Eq. (\ref{eq:meanD}), and quantify how a density staircase modifies the dissipation over a fully convective medium.

\subsubsection{Dependence on the aspect ratio}
First, we {focus on a structure containing} one stably stratified interface, {and} we vary its size. {We do this} by varying the aspect ratio $\varepsilon$, defined by Eq. (\ref{eq:eps}). {It is unclear what the aspect ratio should be in planetary interiors, though {it is likely to be very small}. We note that \citet{LeconteChabrier2012} found that the condition $\varepsilon<1$ was necessary in order for the layered structure to be stable. Here, we explore values ranging from 0.1 to 1. We note that, as} we decrease $\varepsilon$, {the amplitude of the buoyancy bump ($N_0$)} must increase according to Eq. (\ref{eq:amplitudeBVfreq}) {in order} to keep the mean buoyancy frequency $\bar{N} = 10\times (2\Omega)$, constant over the domain.

{The corresponding frequency-averaged quantities can be seen on the left panel of Fig. \ref{fig:nu-5_Eps} for periodic boundary conditions, and of Fig. \ref{fig:nu-5_Eps_ref} for rigid boundary conditions. The results of these calculations (represented by orange filled circles for the total dissipation, and blue squares and red triangles for the viscous and thermal dissipation rates) have been normalised by the frequency-averaged dissipation rate of the fully convective case, which is also indicated by the horizontal blue dashed lines.}

{The qualitative behaviour obtained with both sets of boundary conditions is very similar and is as follows.}
{The frequency-averaged viscous (blue triangles) and thermal (red squares) dissipation rates, do not show a clear monotonic trend. The total frequency-averaged dissipation is approximately constant for $\varepsilon < 0.6$, and is apparently independent of $\varepsilon$ in the astrophysically-relevant case in which $\varepsilon\rightarrow 0$, but it exhibits oscillatory behaviour (with a tendency for larger dissipation) for larger aspect ratios.}
{When comparing the layered case to the fully convective case, our main conclusion is that the frequency-averaged dissipation is larger including layered semi-convection, for all aspect ratios considered here. This is because a layered structure permits more resonances than a fully convective medium, and these contribute to increasing the resulting dissipation. In particular, the additional resonances with short wavelength gravito-inertial, and free gravity modes (which are specific to the layered structure), contribute to the frequency-averaged dissipation to a greater extent than the modes in the sub-inertial range, and the latter are those which are preferably excited in a fully convective medium.}

Finally, {we intuitively expect that} the viscous and thermal dissipation rates in the layered case may be larger when $\nu$ and $\kappa$ are decreased to values that are more relevant to planetary interiors, since more modes are then available to be resonantly excited. {However, we would expect the shortest wavelength modes to be excited less efficiently, so the dominant contribution to the dissipation is probably from the global modes that are excited even as $\nu$ and $\kappa$ are decreased.}

\subsubsection{Dependence on the number of steps $m$}\label{sec:ParameterSpace/Nsteps}
{We now vary the number of stably stratified layers in the domain, $m$. {We recall that this} has {also} the effect of changing the ratio between the characteristic scale of the forcing and the size of the steps. {It is unclear what number of steps could exist in an extended region of layered semi-convection in deep planetary interiors. \cite{LeconteChabrier2012} estimate that it could lie anywhere in a range $10^2 \lesssim m \lesssim 10^9$, but \textit{ab initio} modeling of layered semi-convection in the context of planetary evolution models is very challenging \citep[e.g.][]{VazanEtal2016,GuillotEtal2018}. Here, we explore values from 1 to 5 because calculations with much larger $m$ are computationally very demanding.} As we increase $m$, we keep the aspect ratio constant, $\varepsilon = 0.5$. As a result, the relative size of the layers to the size of the domain is decreasing in order to keep the mean stratification $\bar{N}$ constant and equal to $10 \times (2\Omega)$.}

{The corresponding frequency-averaged quantities can be seen on the middle panel of Fig. \ref{fig:nu-5_Eps} for periodic boundary conditions, and of Fig. \ref{fig:nu-5_Eps_ref} for rigid boundary conditions. The results of these calculations (represented by orange filled circles) have been normalised by the frequency-averaged dissipation rate of the fully convective case, which is also indicated by the horizontal blue dashed lines.} {Our main conclusion is that a region of layered semi-convection is more dissipative (in the frequency-averaged sense) than a fully connvective medium, for the range of $m$ that we have considered here.}

\subsubsection{Dependence on the ratio $\bar{N}/2\Omega$}\label{sec:ParameterSpace/Nmean}
{Finally, we focus on the influence of the parameter $\bar{S}_{\Omega} = \bar{N}/2\Omega$, which characterises the relative strength of the buoyancy force to the Coriolis force. What value this parameter could take in the deep interiors of giant planets is unknown. Here, we vary it from $10^{-2}$ (weakly stratified and/or fast rotator) to $10^2$ (strongly stratified and/or slow rotator) by varying the mean buoyancy frequency $\bar{N}$, and thus the amplitude $N_0$ through the relation given by Eq. (\ref{eq:amplitudeBVfreq}). This is done for one stably stratified interface with an aspect ratio $\varepsilon = 0.2$ and diffusivity coefficients chosen such that $\text{E}=K=10^{-5}$.}

{The corresponding frequency-averaged quantities can be seen on the right panel of Fig. \ref{fig:nu-5_Eps} for periodic boundary conditions, and of Fig. \ref{fig:nu-5_Eps_ref} for rigid boundary conditions. The results of these calculations (represented by orange filled circles) have been normalised by the frequency-averaged dissipation rate of the fully convective case, which is also indicated by the horizontal blue dashed lines.}

{The qualitative behaviour obtained with both sets of boundary conditions is very similar and is as follows.} {Both viscous and thermal frequency-averaged dissipation rates show a clear increase when $\bar{N}>2\Omega$. This can be explained by noting that when $\bar{N}$ increases (and thus $N_0$ increases as we can see from Eq. (\ref{eq:amplitudeBVfreq})), the resonances with short wavelength gravito-inertial waves and free gravity modes of the staircase get stronger.}

{We recall that the free gravity mode given by Eq. (\ref{eq:BQFdisprel_finite_NoRotation}) is such that $\omega$ is proportional to $\bar{N}$ (by some factor depending on wave number and the size of the steps). Thus, the corresponding resonant peak, that is potentially the main contribution to the dissipation (see discussion in Section \ref{subsec:Discussion_consequences}), moves to higher frequency  linearly with $\bar{N}$. If $\bar{N}/2\Omega$ is less than 1, this resonant peaks fall into the sub-inertial frequency range, and the associated dissipation is weak compared to the ones associated with the resonances with short wavelength inertial modes. In this case, considering the medium as fully convective becomes a good approximation to quantify the dissipation. In a giant planet, we expect that this could be the case if the equilibrium layered structure has very large convective steps (of the order of a scale height), and/or if the planet is rotating very rapidly. In this sense, the parameter $\bar{N}/2\Omega$ is the control parameter that determines the relative importance of the effects of the layered structure compared to a fully convective medium, with a layered structure being more dissipative when this ratio is greater than one.}

\section{Conclusions, Discussion \& Future Work}\label{sec:Discussion}
A region of layered semi-convection consists of density staircases, in which convective layers are separated by thin stably stratified interfaces. We have presented exploratory linear calculations to study how layered semi-convection, that is potentially present in giant planet interiors, affects the rates of tidal dissipation. We adopted a local Cartesian model to study the dissipation of short-wavelength internal (gravito-inertial) waves excited in a region of layered semi-convection by a gravitational tidal-like forcing. {We have computed the response of such a density structure, and we have provided a physical interpretation for each of the associated resonances with enhanced dissipation. We also computed frequency-averaged dissipation rates to determine how dissipative a semi-convective medium {compares} to a fully convective medium as the various parameters of our model are varied. Our calculations were undertaken with two different and complementary sets of boundary conditions{. First, vertically periodic ones, relevant to study a portion of a vertically extended region of layered semi-convection. Second, vertically rigid (and stress-free) boundary conditions, equivalent to a plane-parallel model of a region filled with layered semi-convection. This was done assuming horizontal periodicity in each case, and the results lead to modifications of resulting global modes.}}

Our primary intended application is to understand tidal dissipation in giant planets and the consequent evolution of their natural satellite systems, though our results could also be relevant for tides in short-period extra-solar planets interacting with their host stars. Our main conclusions are the following:

\begin{itemize}
\item A region of layered semi-convection possesses a richer set of free modes than a fully convective medium  -- which is the model that is usually adopted for giant planet deep interiors. As a result, more resonances can potentially be excited compared to a convective medium. This makes it more likely for a satellite {or host star} to enter a resonance with enhanced tidal dissipation, potentially by several orders of magnitude.\\[-6pt]
\item These resonances are more broadly distributed over the frequency spectrum compared to a fully convective medium. Short wavelength inertial modes can be excited (like in a fully convective medium), but we have also identified short wavelength gravito-inertial modes (localised within the stably-stratified interfaces, though we expect these to be very thin in reality), and gravity modes (g-modes), influenced by the mean stratification, characterised by the mean buoyancy frequency $\bar{N}$. Thus, the frequency range in which we expect resonances to be excited is extended from $-2\Omega \lesssim \omega \lesssim 2\Omega$ in a convective medium, to $-\bar{N} \lesssim \omega \lesssim \bar{N}$ in a medium with layered semi-convection.\\[-6pt]
\item Extrapolating the trends that we have observed, additional resonances are expected for smaller diffusivities and thinner interfaces, potentially leading to an enhancement in the rates of tidal dissipation over an even wider frequency range in the astrophysical regime.
\end{itemize}

\subsection{Consequences for tidal dissipation in giant planets}\label{subsec:Discussion_consequences}
Our calculations within this local Cartesian model cannot be used to directly make \textit{quantitative} predictions for the rates of tidal dissipation in giant planets. This is because the forcing that we have adopted is highly idealised. Nevertheless, we have clearly identified some of the most important physical effects that may lead to higher dissipation in a region of layered semi-convection compared to a fully convective medium.

We have seen that the dissipation is higher for a wider range of frequencies in the layered case compared with a fully convective medium in our model. This corresponds with the excitation of particular global mode{s} that {are} resonant with our adopted forcing. 
{Adopting a realistic dependence for the forcing amplitude on the tidal frequency, we have seen that the \textit{frequency-averaged} dissipation rates exhibited a clear trend to higher dissipation in the layered case versus the fully convective case. This is because the primary contribution to the averaged dissipation is due to the resonances with free gravity modes.}

{In reality, the scale over which the equilibrium tide (which acts as an effective forcing for tidal waves) varies should be much larger than the size of the box that we model, and it should be radially node-less. For clarity (and numerical convenience), we assumed in this study the forcing to be periodic in the local radial direction, which has the effect of artificially enhancing the strength of this resonance with the box-scale inertial mode if the medium is fully convective, but only when periodic boundary conditions are assumed. This is indeed a drawback from our local analysis, since in the realistic tidal problem the forcing should vary on a large length-scale compared to the size of the semi-convective layers. When considering the realistic tidal problem, a possibility is that the resonances with short wavelength inertial and gravito-inertial modes (described in Section \ref{sec:modes}) could be significantly weakened, because the coupling between the large-scale tidal forcing and the small-scale oscillation modes could be decreased. However, we expect that the astrophysically relevant smaller diffusivities would partly counteract this effect, since less damping of the resonances would be expected. {On the other hand, the resonances with the free gravity modes of the staircase (which have internal gravity wave-like character with respect to the mean stratification) should remain robust and will probably be the most significant contribution to the dissipation in layered semi-convection regions.}}

{In our model we also considered the effect of adopting rigid boundary conditions in the vertical direction, which may be a closer approximation to the global problem in which an extended envelope would be filled with layered semi-convection. We found that this had the effect of modifying the resonances with free gravity-modes of the staircase, which were then more numerous when additional layers were included. This was consistent with the work of \citet{BelyaevQuataertFuller2015} and \citetalias{ABM2017}. Our other main conclusions were independent of the boundary conditions that we adopted in the vertical direction.}

While this paper constitutes a necessary first step, in order to compute astrophysically and \textit{quantitatively} meaningful tidal dissipation rates as a function of the structural parameters of the region of layered semi-convection, we crucially need to extend this study to spherical geometry. This will enable us to consider a more realistic situation where curvature effects are included, and the tidal gravitational forcing can be included self-consistently. 
Performing a calculation along these lines in spherical geometry will be the focus of a future paper.

In the near future, as we hope that our understanding of giant planet internal structures will be improved (especially thanks to the Juno spacecraft), it is important to be able to include the most important structural details and their effects on the rates of tidal dissipation. We recall that, based on astrometric observations spanning more than a century, tidal dissipation in Jupiter and Saturn has been found to be much higher than previously thought \citep[e.g.][]{LaineyEtal2017}. These results have not yet been {fully} explained, motivating us to consider more sophisticated tidal models. {However, we note that several mechanisms have already been proposed to explain high tidal dissipation rates in giant planets. \citet{OgilvieLin2004} have studied in detail the frequency-dependence of the dissipation of tidal inertial waves by turbulent friction in convective envelopes \citep[see also][]{AuclairDesrotourEtal2015,MathisEtal2016}. Moreover, \citet{RemusMathisZahn2012,RemusEtal2015} have studied the visco-elastic dissipation of a possible rocky/icy core in central regions of giant planets, for different rheologies and tidal frequencies. It has then been shown by \citet{GuenelMathisRemus2014} that these two potential sources of tidal dissipation could be of comparable strengths depending on the rheology of the core, {while the} {viscous friction in a solid core was found to be compatible with the tidal dissipation rates and frequency-dependence found by \citet{LaineyEtal2017} in Saturn.} In particular, these recent works point out very clearly the importance of being able to carefully take into account each type of dissipation mechanism. To do so, it is in turn crucial to rely on realistic models of giant planet interiors. Finally, \citet{FullerEtal2016} have studied the possibility that some moons of Jupiter or Saturn could migrate outwards while being locked in a resonance (for which the dissipation is very efficient) with a tidal gravito-inertial mode of the central giant planet, leading to fast outward migration.}

This paper thus represents a first step towards understanding how layered semi-convection, that is potentially present in giant planets, affects the rates of tidal dissipation. We tentatively conclude that layered semi-convection, if present in giant planets, could play an important role in enhancing the rates of tidal dissipation in giant planets, {alongside the mechanisms that have been described in the previous paragraph. These} may be key to explain the high tidal dissipation rates observed by \cite{LaineyEtal2009,LaineyEtal2012,LaineyEtal2017,PolycarpeEtal2018} in Jupiter and Saturn, {and could be important in the evolution of short-period extrasolar planetary systems}.

\subsection{Link between transmission and dissipation}
In \citetalias{ABM2017}, we showed that short wavelength internal (and inertial) waves are only efficiently transmitted across a region of layered semi-convection if they are resonant with a free mode of the staircase. Here, we showed that free modes of the staircase can be excited by a gravitational tidal-like forcing, leading to enhanced dissipation. It is also interesting to note that a region of layered semi-convection lying below a convective region could act as a rigid wall for waves that do not excite one of the free modes. Those waves would then only dissipate their energy in the overlying convective region. However, we expect tidal waves that excite a free mode of the staircase to be efficiently transmitted, and thus to be able to reach the deepest layers of the fluid envelope, where they may be dissipated and could heat the deep interior.

\subsection{Prospects and future work}
Further work is required to explore and confirm the influence of layered semi-convection on tidal dissipation in global models. Other mechanisms should also be taken into account, such as differential rotation \citep{BaruteauRieutord2013,GuenelEtal2016a,GuenelEtal2016b}, magnetic fields \citep{BarkerLithwick2014,Wei2016,LinOgilvie2018,Wei2018}{, and the impact of rotation on the semiconvective background structure \citep{MollGaraud2017}}.

{In the context of the ongoing Juno mission, these two physical ingredients have recently been shown to be of significant importance in Jupiter. Indeed, new constraints obtained on Jupiter's zonal flows suggest that differential rotation is significant in a shell extending down to 3,000 km in radius. The deeper interior is expected to rotate as a solid body due to the action of magnetic stresses associated with an increase of the electric conductivity of the gas at this location \citep{GuillotEtal2018Nature}. In the near future, the Juno mission should provide additional constraints on the internal structure of Jupiter{, and its magnetic field \citep{ConnerneyEtal2018}. These} could hopefully be extrapolated to further constrain other giant planets (including hot Jupiters) and brown dwarfs. Including additional physical effects, motivated by these observational constraints, is currently our best way to build more realistic tidal models of systems involving such bodies.}

\begin{acknowledgements}
The authors thank the referee for a constructive report that allowed us to improve the clarity and completeness of this paper. QA thanks Kyle Augustson for useful discussions. QA and SM acknowledge funding by the European Research Council through ERC SPIRE grant 647383, the support of ISSI to the international team Encelade 2.0, and the PLATO CNES grant at CEA-Saclay. AJB was supported by the Leverhulme Trust through the award of an Early Career Fellowship and by STFC Grant ST/R00059X/1. 
\end{acknowledgements}

\begin{appendix}
\section{Derivation of the forced Poincar\'e equation}
\label{app:PoincareEq}
The linearised system we consider, when we adopt the Boussinesq approximation, is given by Eqs. (\ref{momx})--(\ref{ener}). The aim of this appendix is to derive the forced Poincar\'e equation.

First, taking the combination ${\uppartial_y}$(\ref{momz}) $-~{\uppartial_z}$(\ref{momy}) gives
\begin{equation}
D_{\nu}\left(\frac{\uppartial w}{\uppartial y}-\frac{\uppartial v}{\uppartial z}\right) - (\bm{f}\cdot\bm{\nabla})u
= \frac{\uppartial b}{\uppartial y} + [\bm{\nabla}\times\bm{F}]_x,
\label{eq:1stcomb}
\end{equation}
where $[\bm{\nabla}\times\bm{F}]_x \equiv (\bm{\nabla}\times\bm{F})\cdot\hat{\bm{e}}_x = {\uppartial_y}F_z - {\uppartial_z}F_y$, and $\bm{f} = \left(0,\tilde{f},f\right)$.
Then, taking the combination ${\uppartial_z}$(\ref{momx}) $-~{\uppartial_x}$(\ref{momz}) and using Eq. (\ref{cont}) gives
\begin{equation}
D_{\nu}\left(\frac{\uppartial u}{\uppartial z}-\frac{\uppartial w}{\uppartial x}\right) - (\bm{f}\cdot\bm{\nabla})v
= - \frac{\uppartial b}{\uppartial x} + [\bm{\nabla}\times\bm{F}]_y,
\label{eq:2ndcomb}
\end{equation}
where $[\bm{\nabla}\times\bm{F}]_y \equiv (\bm{\nabla}\times\bm{F})\cdot\hat{\bm{e}}_y = {\uppartial_z F_x} - {\uppartial_x F_z}$.\\
Then, taking the combination ${\uppartial_x}$(\ref{momy}) $-~{\uppartial_y}$(\ref{momx}) and using Eq. (\ref{cont}) gives
\begin{equation}
D_{\nu}\left(\frac{\uppartial v}{\uppartial x}-\frac{\uppartial u}{\uppartial y}\right) - (\bm{f}\cdot\bm{\nabla})w
= [\bm{\nabla}\times\bm{F}]_z,
\label{eq:3rdcomb}
\end{equation}
where $[\bm{\nabla}\times\bm{F}]_z \equiv (\bm{\nabla}\times\bm{F})\cdot\hat{\bm{e}}_z = {\uppartial_x F_y} - {\uppartial_y F_x}$.\\
Then, by taking the combination $D_{\kappa}D_{\nu}\left({\uppartial_y}\text{(\ref{eq:1stcomb})} - {\uppartial_x}\text{(\ref{eq:2ndcomb})}\right)$ and using equations (\ref{cont}), (\ref{ener}) and (\ref{eq:3rdcomb}), we finally obtain an equation for the vertical component of the velocity $w$,
\begin{equation}
D_{\kappa}D_{\nu}^2 \nabla^2 w + D_{\kappa}(\bm{f} \cdot \bm{\nabla})^2 w + D_{\nu}\left[N^2 \nabla_{\perp}^2\right] w
= \bm{\mathcal{O}} \cdot (\bm{\nabla} \times \bm{F}),
\label{eq:forced_Poincare}
\end{equation}
where $\nabla_{\perp}^2 \equiv {\uppartial_{xx}} + {\uppartial_{yy}}$, $D_{\alpha} = \uppartial_t - \alpha\nabla^2$, and
\begin{equation}
\bm{\mathcal{O}} \equiv D_{\kappa}
\left(
\begin{array}{c}
      D_{\nu}{\uppartial_y}\\
      -D_{\nu}{\uppartial_x}\\
      -\bm{f}\cdot\bm{\nabla}
\end{array}
\right).
\label{operator}
\end{equation}

\end{appendix}

\bibliographystyle{aa} 
\bibliography{staircase} 

\end{document}